\documentclass[journal]{IEEEtran}

\ifCLASSINFOpdf
\else
\fi
\usepackage{algpseudocode}
\usepackage{algorithm}
\usepackage[dvips]{graphicx}
\usepackage{latexsym}
\usepackage{amssymb}
\usepackage{amsmath}
\usepackage{multirow}
\usepackage{xcolor}
\usepackage{lipsum}
\usepackage{multicol}
\usepackage{enumerate}
\usepackage{algorithm}
\usepackage{algorithmicx}
\usepackage{algpseudocode}
\usepackage{amsmath}
\usepackage{graphicx}
\usepackage{subfig}
\usepackage[utf8]{inputenc}
\usepackage{float}

\begin{document}
%
\title{Interference Exploitation Precoding Made Practical: Closed-Form Solutions with Optimal Performance}
%
%
%

\author{Ang~Li,~\IEEEmembership{Student Member,~IEEE}, and Christos Masouros,~\IEEEmembership{Senior Member,~IEEE}

\thanks{Manuscript received TIME; revised TIME.}
\thanks{A. Li and C. Masouros are with the Department of Electronic and Electrical Engineering, University College London, Torrington Place, London, WC1E 7JE, UK (e-mail: chris.masouros@ieee.org, ang.li.14@ucl.ac.uk).}
\thanks{This work was supported by the Royal Academy of Engineering, UK, the Engineering and Physical Sciences Research Council (EPSRC) project EP/M014150/1 and the China Scholarship Council (CSC).}
}



\maketitle

\begin{abstract}
In this paper, we propose closed-form precoding schemes with optimal performance for constructive interference (CI) exploitation in the multiuser multiple-input single-output (MU-MISO) downlink. We first consider an optimization where we maximize the distance between the constructive region and the detection thresholds. The cases of both strict and non-strict phase rotation are considered and can further be formulated as convex optimization problems. For optimization with strict phase rotation, we mathematically derive the optimal beamforming structure with Lagrangian and Karush-Kuhn-Tucker (KKT) conditions. By formulating its dual problem, the optimization problem is further shown to be equivalent to a quadratic programming (QP) over a simplex, which can be solved more efficiently. We then extend our analysis to the case of non-strict phase rotation, where it is mathematically shown that a $K$-dimensional optimization for non-strict phase rotation is equivalent to a $2K$-dimensional optimization for strict phase rotation in terms of the problem formulation. The connection with the conventional zero-forcing (ZF) precoding is also discussed. Based on the above analysis, we further propose an iterative closed-form scheme to obtain the optimal beamforming matrix, where within each iteration a closed-form solution can be obtained. Numerical results validate our analysis and the optimality of the proposed iterative scheme, and further show that the proposed closed-form scheme is more efficient than the conventional QP algorithms with interior-point methods, which motivates the use of CI beamforming in practical wireless systems.
\end{abstract}

\begin{IEEEkeywords}
MIMO, transmit beamforming, constructive interference, optimization, Lagrangian, closed-form solutions.
\end{IEEEkeywords}

%
\IEEEpeerreviewmaketitle

\section{Introduction}
%
%
%
%
\IEEEPARstart{M}{ULTIPLE}-input multiple-output (MIMO) systems have been widely acknowledged as a promising technology in the field of wireless communications, due to the significant gains over single-antenna systems \cite{r1}. When the channel knowledge is known at the base station (BS), the capacity-achieving dirty-paper coding (DPC) scheme is proposed in \cite{r2} by pre-subtracting the interference prior to transmission. However, DPC is difficult to implement in practical systems due to the impractical assumption of infinite length of codewords and its high computational cost. To achieve a compromise between performance and complexity, its non-linear counterparts in the form of Tomlinson-Harashima precoding (THP) \cite{r3} and vector perturbation (VP) \cite{r4} have been proposed, which however are still too complicated for practice due to the inclusion of the sophisticated sphere-search algorithms. Therefore, low-complexity linear precoding schemes based on zero-forcing (ZF) have received increasing research attention \cite{r5}, and a regularized ZF (RZF) scheme is proposed in \cite{r6} to further improve the performance of ZF. On the other hand, transmit beamforming schemes based on optimization have also been a popular research topic \cite{r7}\nocite{r8}\nocite{r9}\nocite{r10}\nocite{r11}\nocite{r12}-\cite{r13}. Among the optimization-based schemes, one form of the optimization known as signal-to-interference-plus-noise ratio (SINR) balancing is to maximize the minimum SINR subject to a total power constraint \cite{r7}, \cite{r8} or a per-antenna power constraint \cite{r9}. An alternative downlink beamforming targets at minimizing the total transmit power at the BS subject to a minimum SINR requirement \cite{r10}-\cite{r12}. It has been shown that the power minimization problems can be formulated either as a virtual uplink problem with power control or as a semi-definite programming (SDP) and solved via semi-definite relaxation (SDR) \cite{r11}. As for the SINR balancing problem, it is proven to be an inverse problem to the power minimization optimization, based on which schemes via bisection search \cite{r7} and iterative algorithms \cite{r10} have been proposed.

Nevertheless, both the above precoding schemes and the optimization-based transmit beamforming designs mentioned above have ignored the fact that interference can be beneficial and further exploited on an instantaneous basis \cite{r14}, \cite{r15}. The concept of constructive interference (CI) was firstly introduced in \cite{r30}, where it is shown that the instantaneous interference can be categorized into constructive and destructive. A modified ZF precoding scheme is then proposed in \cite{r16}, where the constructive interference is exploited while the destructive interference is cancelled. A correlation rotation scheme has been further proposed in \cite{r17}, where it is shown that the destructive interference can be manipulated and rotated such that all the interference becomes constructive. Symbol-level transmit beamforming schemes based on convex optimization for CI has been proposed in \cite{r18}, \cite{r19}, where the concept of constructive region is introduced to relax the strict phase rotation constraint in \cite{r17} and achieve an improved performance. Further studies on the optimization-based CI beamforming schemes can be found in \cite{r19}\nocite{r20}\nocite{r31}-\cite{r21}. Due to the performance benefits over conventional schemes, the concept of CI has been extended to many wireless application scenarios, including cognitive radio \cite{r22}, \cite{r23}, constant envelope precoding \cite{r24}, wireless information and power transfer \cite{r25} and mutual coupling exploitation \cite{r26}. The above studies show that MIMO systems can benefit from the CI with a symbol-level beamforming. Nevertheless, while the performance of CI-based beamforming approaches is superior, they need to solve a convex optimization problem, which can be computationally inefficient, especially when executed on a symbol-by-symbol basis.

In this paper, we design low-complexity optimal and sub-optimal solutions for CI precoding, culminating in closed-form iterative precoders. We consider an optimization problem where we maximize the distance between the constructive region and the detection thresholds such that the effect of CI is maximized. We firstly consider the optimization for strict phase rotation, where the phases of the interfering signals are rotated such that they are strictly aligned to the symbol of interest. By analyzing the formulated second-order cone programming (SOCP) optimization with Lagrangian and KKT conditions, we derive the structure of the optimal beamforming matrix, which leads to an equivalent optimization and further simplifies the beamforming design. By formulating the dual problem of the equivalent optimization problem, it is mathematically shown that the optimization for CI beamforming is equivalent to a quadratic programming (QP) optimization over a simplex, which finally leads to a closed-form expression. We extend our analysis to the case of non-strict phase rotation, where the phases of the interfering signals are rotated such that the resulting interfered signal is located within the constructive region. By following a similar approach for the case of strict phase rotation, we analytically show that the optimal beamforming matrix for theses two scenarios shares a similar closed-form expression, and a $K$-dimensional optimization for non-strict phase rotation is equivalent to a $2K$-dimensional optimization for strict phase rotation in terms of the problem formulation. Our above analysis also provides some insights on the connection between the CI beamforming and the generic ZF precoding. 

Moreover, our efforts to facilitate the symbol-level CI beamforming in practice culminate in an iterative closed-form scheme to efficiently obtain the optimal beamforming matrix, where a closed-form solution is obtained within each iteration. We show that only in a few iterations, the closed-form approach obtains optimal performance. Numerical results validate our above analysis and the optimality of the proposed iterative closed-form method for both strict and non-strict phase rotation. Moreover, it is numerically shown that the proposed iterative approach is more time-efficient compared to the conventional QP algorithms based on interior-point methods, especially when the number of users is small. By constraining the maximum number of iterations, we further obtain a flexible performance-complexity tradeoff for the proposed iterative method, based on its connection with conventional ZF precoding. Both of the above motivate the use of CI-based beamforming in practical wireless systems.

For reasons of clarity, we summarize the contributions of this paper as:
\begin{enumerate}

\item We formulate the optimization problem for CI-based beamforming, where we maximize the distance between the constructive region and the detection thresholds. We derive the optimal beamforming matrix for strict phase rotation and further formulate an equivalent and simplified optimization problem.

\item The optimization for strict phase rotation is transformed and further shown to be equivalent to a QP problem over a simplex, which can be more efficiently solved than the originally formulated problem.

\item We extend our analysis to the case of non-strict phase rotation, where the closed-form expression is similar to the case of strict phase rotation. It is further shown that a $K$-dimensional optimization for non-strict phase rotation is equivalent to a $2K$-dimensional optimization for strict phase rotation in terms of the problem formulation.

\item We analytically study the connection between the CI beamforming and the ZF precoding, where it is shown that ZF precoding can be regarded as a special case of CI-based beamforming with all the dual variables being zero.

\item We further propose an iterative closed-form scheme to obtain the optimal beamforming matrix for both the strict and non-strict phase rotation cases, where within each iteration a closed-form solution can be derived. We show that the closed-form precoder obtains an optimal performance in only a few iterations.

\end{enumerate}

The remainder of this paper is organized as follows. Section II introduces the system model and briefly reviews CI. Section III includes the analysis for the optimization problems with both strict and non-strict phase rotation constraints. The connection between the CI beamforming and conventional ZF precoding is discussed in Section IV. The proposed iterative closed-form scheme is introduced in Section V. The numerical results are shown in Section VI, and Section VII concludes the paper.

$Notations$: $a$, $\bf a$, and $\bf A$ denote scalar, vector and matrix, respectively. ${( \cdot )^T}$, ${( \cdot )^H}$ and $tr\left\{  \cdot  \right\}$ denote transposition, conjugate transposition and trace of a matrix, respectively. $j$ denotes the imaginary unit, and ${\rm{vec}}\left(  \cdot  \right)$ denotes the vectorization operation. ${\bf{A}}\left( {k,i} \right)$ denotes the entry in the $k$-row and $i$-th column of $\bf A$. $\left|  \cdot  \right|$ denotes the absolute value of a real number or the modulus of a complex number, and $\left\|  \cdot  \right\|_F$ denotes the Frobenius norm. ${{\cal C}^{n \times n}}$ represents an $n \times n$ matrix in the complex set, and $\bf I$ denotes the identity matrix. $\Re ( \cdot )$ and $\Im ( \cdot )$ denote the real and imaginary part of a complex number, respectively. $card\left(  \cdot  \right)$ denotes the cardinality of a set.

\section{System Model and Constructive Interference}
In this section, the system model that we consider is firstly introduced, followed by a brief review of CI and the constructive region.

\begin{figure*}
\centering
\includegraphics[scale=0.4]{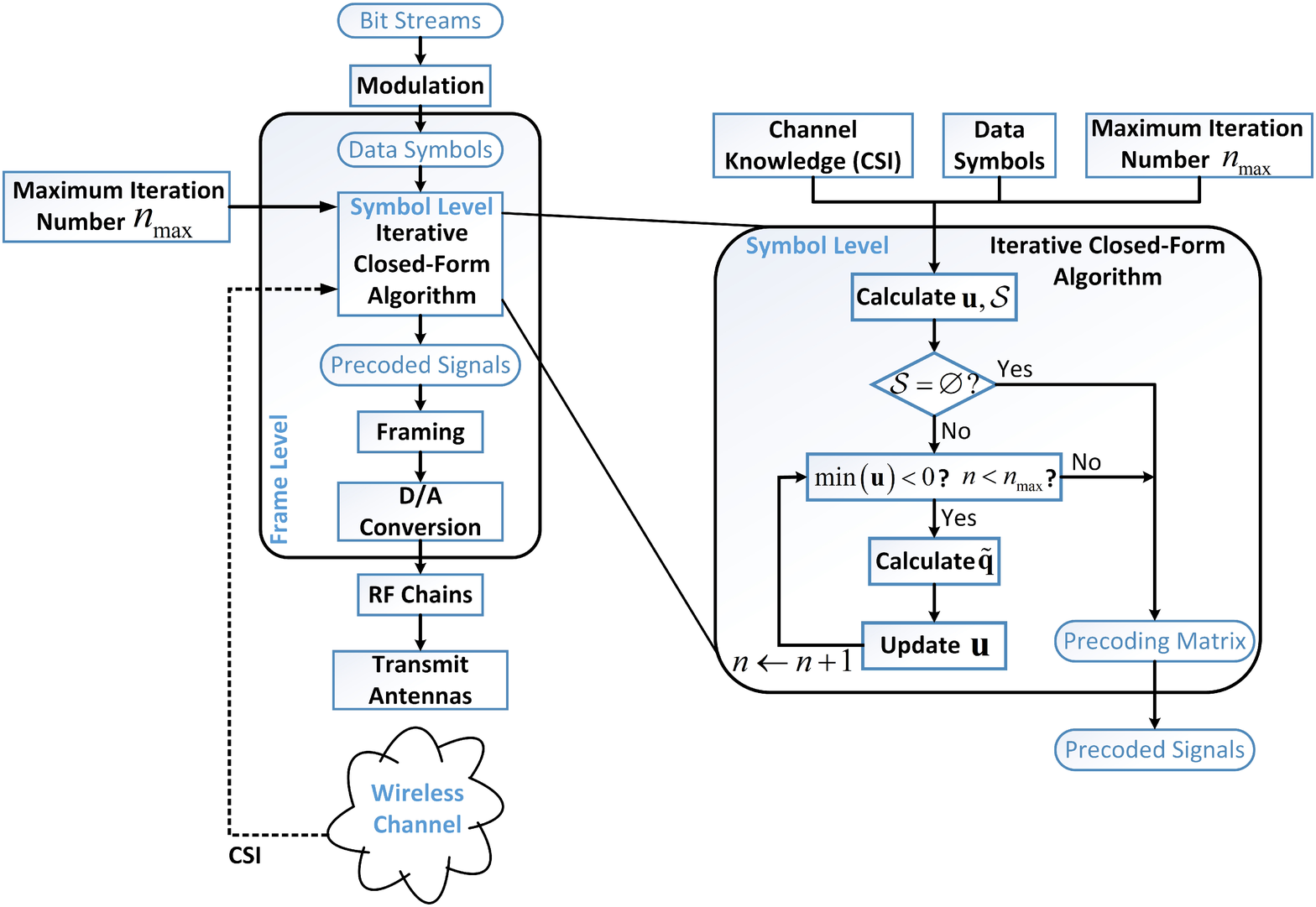}
\caption{A block diagram for the proposed symbol level precoding based on constructive interference}
\end{figure*}

\subsection{System Model}
We consider a multiuser MISO system in the downlink, where the BS structure with a symbol-level precoding is depicted in Fig.~1, where the iterative closed-form algorithm will be introduced in Section V. The BS with $N_t$ transmit antennas is simultaneously communicating with $K$ single-antenna users in the same time-frequency resource, where $K \le N_t$. We focus on the transmit beamforming designs and perfect CSI is assumed throughout the paper. The data symbol vector is assumed to be from a normalized PSK modulation constellation \cite{r18}, denoted as ${\bf s} \in {\cal C}^{K \times 1}$. Then, the received signal at the $k$-th user can be expressed as
\begin{equation}
r_k= {\bf h}_k{\bf Ws} + n_k,
\label{eq_1}
\end{equation}
where ${\bf h}_k \in {\cal C}^{1 \times N_t}$ denotes the flat-fading Rayleigh channel vector from user $k$ to the BS, and each entry in ${\bf h}_k$ follows a standard complex Gaussian distribution. ${\bf W} \in {\cal C}^{N_t \times K}$ is the beamforming matrix and $n_k$ is the additive Gaussian noise with zero mean and variance $\sigma^2$ at the receiver.

\begin{figure}[!t]
\centering
\includegraphics[scale=0.4]{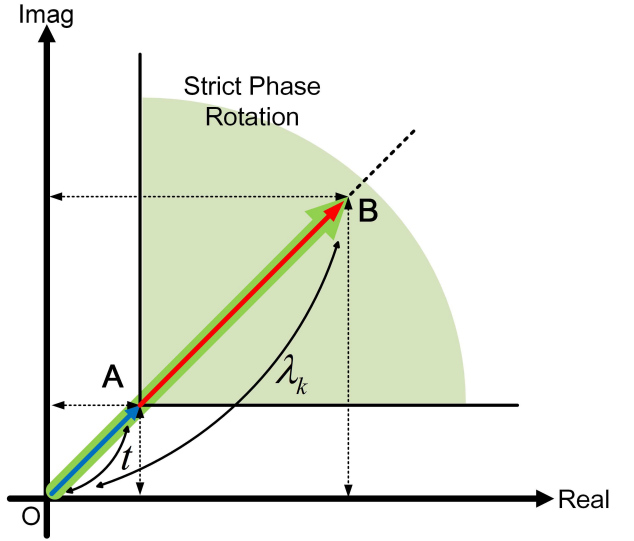}
\caption{Constructive interference, QPSK, strict phase rotation}
\end{figure}

\subsection{Constructive Interference}
CI is defined as the interference that pushes the received signals away from the detection thresholds \cite{r14}-\cite{r17}. CI for strict phase rotation refers to the cases where the phases of the interfering signals are controlled and rotated, such that they are strictly aligned to those of the data symbols of interest \cite{r17}. The constructive region has been further introduced in \cite{r18}, where it is shown that the phases of the interfering signals may not be necessarily strictly aligned to that of the data symbols of interest, known as the non-strict phase rotation. It is demonstrated that, as long as the resulting interfered signals are located in the constructive region, this increases the distance to the detection thresholds and returns an improved performance. To show this intuitively, in Fig. 2 and Fig. 3 we depict the case for strict phase rotation and non-strict phase rotation respectively, where the constellation point $\left( {\frac{1}{{\sqrt 2 }} + \frac{1}{{\sqrt 2 }} \cdot j} \right)$ from a normalized QPSK constellation is employed as the example to illustrate these two cases. We can observe that for both strict phase rotation and non-strict phase rotation, the distance of the received signals to the detection thresholds is increased, which will improve the detection performance.

\section{Constructive Interference Beamforming}
In this section, we firstly focus on the CI beamforming for strict phase rotation, and we further extend our analysis to the case of non-strict phase rotation. 

\begin{figure}[!t]
\centering
\includegraphics[scale=0.4]{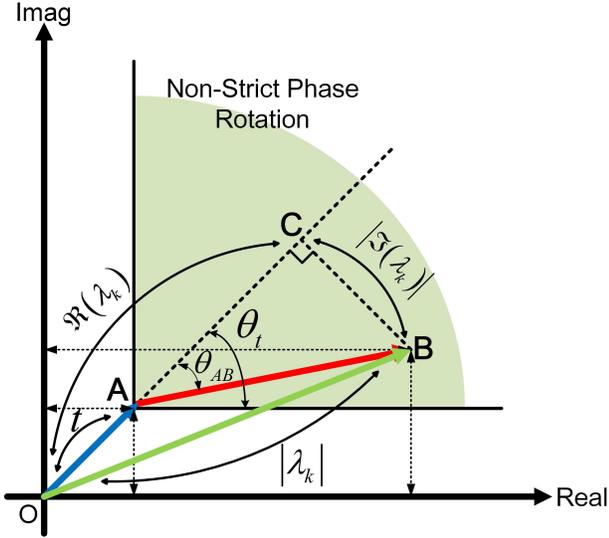}
\caption{Constructive region, QPSK, non-strict phase rotation}
\end{figure} 

\subsection{Strict Phase Rotation}
Before formulating the optimization problem, based on the geometry of the modulation constellation we firstly construct the conditions that the beamformer should satisfy to achieve the strict phase rotation. In Fig. 2, without loss of generality we denote $\mathop {OA}\limits^ \to   = t \cdot {s_k}$ and $t = |\mathop {OA}\limits^ \to  |$ is the object to be maximized. We further assume that the node `B' denotes the noiseless received signal for user $k$ that is co-linear to $\mathop {OA}\limits^ \to$ for strict phase rotation, which leads to
\begin{equation}
\mathop {OB}\limits^ \to   = {{\bf{h}}_k}{\bf{Ws}}.
\label{eq_2}
\end{equation} 
Then, by introducing a real-valued scaling factor $\lambda_k$, we further express $\mathop {OB}\limits^ \to$ as
\begin{equation}
\mathop {OB}\limits^ \to = {{\bf{h}}_k}{\bf{Ws}} = {\lambda _k}{s_k},
\label{eq_3}
\end{equation}
where based on the geometry we can obtain that ${\lambda _k}$ is a real number, and the condition on ${\lambda _k}$ to achieve CI for strict phase rotation is given by
\begin{equation}
\lambda_k \ge t, {\kern 3pt} \forall k \in {\cal K},
\label{eq_4}
\end{equation} 
where ${\cal K} = \left\{ {1,2, \cdots ,K} \right\}$. With the above formulation, we can construct the optimization problem for strict phase rotation as
\begin{equation}
\begin{aligned}
&\mathcal{P}_1: {\kern 3pt} \mathop {\max }\limits_{{{\bf{W}}}, {\kern 1pt} t} {\kern 3pt} t \\
&{\kern 0pt} s. t. {\kern 10pt} {{\bf{h}}_k}{\bf Ws} = {\lambda _k}{s_k}, {\kern 3pt} \forall k \in {\cal K} \\
&{\kern 24pt} \lambda_k \ge t, {\kern 3pt} \forall k \in {\cal K}\\
&{\kern 22pt} \left\| {{\bf{Ws}}} \right\|_F^2 \le {p_0}
\label{eq_5}
\end{aligned}
\end{equation}
where $p_0$ denotes the total available transmit power. A symbol-level power constraint is employed, as the exploitation of CI is related to the transmit symbol vector, which will also be shown mathematically in the following. ${\cal P}_1$ belongs to the SOCP and can be solved with convex optimization tools such as CVX \cite{r18}. We decompose the beamforming matrix into vectors
\begin{equation}
{\bf{W}} = \left[ {{{\bf{w}}_1},{{\bf{w}}_2}, \cdots ,{{\bf{w}}_K}} \right],
\label{eq_6}
\end{equation}
and based on the virtual multicast formulation in \cite{r18} we obtain that each ${\bf w}_i s_i$ is identical. This leads to the equivalent transformation of the power constraint, given by
\begin{equation}
\left\| {{\bf{Ws}}} \right\|_F^2 \le {p_0} \Rightarrow \sum\limits_{i = 1}^K {s_i^H{\bf{w}}_i^H{{\bf{w}}_i}{s_i}} \le \frac{p_0}{K}.
\end{equation}
We further transform ${\cal P}_1$ in \eqref{eq_5} into a standard minimization problem, expressed as
\begin{equation}
\begin{aligned}
&\mathcal{P}_2: {\kern 3pt} \mathop {\min }\limits_{{{\bf w}_i}, {\kern 1pt} t} {\kern 3pt} -t \\
&{\kern 0pt} s. t. {\kern 10pt} {{\bf{h}}_k}\sum\limits_{i = 1}^K {{{\bf{w}}_i}{s_i}}  - {\lambda _k}{s_k} = 0, {\kern 3pt} \forall k \in {\cal K} \\
&{\kern 24pt} t - {\lambda _k} \le 0, {\kern 3pt} \forall k \in {\cal K}\\
&{\kern 22pt} \sum\limits_{i = 1}^K {s_i^H{\bf{w}}_i^H{{\bf{w}}_i}{s_i}}  - \frac{p_0}{K} \le 0
\label{eq_7}
\end{aligned}
\end{equation}
In the following we analyze ${\cal P}_2$ with Lagrangian and KKT conditions. The Lagrangian of ${\cal P}_2$ is expressed as \cite{r27}
\begin{equation}
\begin{aligned}
&{\cal L}\left( {{{\bf{w}}_i},t,{\delta _k},{\mu _k},{\mu _0}} \right) =  - t + \sum\limits_{k = 1}^K {{\delta _k}\left( {{{\bf{h}}_k}\sum\limits_{i = 1}^K {{{\bf{w}}_i}{s_i}}  - {\lambda _k}{s_k}} \right)} \\
&+ \sum\limits_{k = 1}^K {{\mu _k}\left( {t - {\lambda _k}} \right)}  + {\mu _0}\left( {\sum\limits_{i = 1}^K {s_i^H{\bf{w}}_i^H{{\bf{w}}_i}{s_i}}  - \frac{p_0}{K}} \right),
\label{eq_8}
\end{aligned}
\end{equation}
where $\delta_k$, $\mu_k$ and $\mu_0$ are the dual variables, and we have $\mu_0 \ge 0$ and $\mu_k \ge 0$, $\forall k \in {\cal K}$. Based on the Lagrangian in \eqref{eq_8}, the KKT conditions for optimality can be obtained as
\begin{IEEEeqnarray}{rCl} 
\IEEEyesnumber
\frac{{\partial {\cal L}}}{{\partial t}} =  - 1 + \sum\limits_{k = 1}^K {{\mu _k}}  = 0 {\kern 30pt} \IEEEyessubnumber* \label{eq_9a} \\
\frac{{\partial {\cal L}}}{{\partial {{\bf{w}}_i}}} = \left( {\sum\limits_{k = 1}^K {{\delta _k} \cdot {{\bf{h}}_k}} } \right){s_i} + {\mu _0} \cdot {\bf{w}}_i^H = {\bf{0}}, {\kern 2pt} \forall i \in {\cal K} {\kern 30pt} \label{eq_9b} \\
{\delta _k}\left( {{{\bf{h}}_k}\sum\limits_{i = 1}^K {{{\bf{w}}_i}{s_i}}  - {\lambda _k}{s_k}} \right) = 0, {\kern 2pt} \forall k \in {\cal K}{\kern 30pt} \label{eq_9c} \\
{\mu _k}\left( {t - {\lambda _k}} \right) = 0, {\kern 2pt} \forall k \in {\cal K} {\kern 30pt} \label{eq_9d}\\
{\mu _0}\left( {\sum\limits_{i = 1}^K {s_i^H{\bf{w}}_i^H{{\bf{w}}_i}{s_i}}  - \frac{p_0}{K}} \right) = 0 {\kern 30pt} \label{eq_9e}
\end{IEEEeqnarray}
Based on \eqref{eq_9b}, it is firstly obtained that $\mu_0 \ne 0$, and with the fact that $\mu_0 \ge 0$ we can further obtain $\mu_0 > 0$. Then, ${\bf w}_i^H$ in \eqref{eq_9b} can be expressed as
\begin{equation}
{\bf{w}}_i^H =  - \frac{1}{{{\mu _0}}}\left( {\sum\limits_{k = 1}^K {{\delta _k} \cdot {{\bf{h}}_k}} } \right){s_i}, {\kern 2pt} \forall i \in {\cal K}.
\label{eq_10}
\end{equation}
By denoting 
\begin{equation}
{\upsilon _k} =  - \frac{{\delta _k^H}}{{{\mu _0}}}, {\kern 2pt} \forall k \in {\cal K},
\label{eq_11}
\end{equation}
where we note that $\delta_k$ can be complex, the expression of ${\bf w}_i$ is obtained as
\begin{equation}
{{\bf{w}}_i} = \left( {\sum\limits_{k = 1}^K {{\upsilon_k} \cdot {\bf{h}}_k^H} } \right)s_i^H, {\kern 2pt} \forall k \in {\cal K}.
\label{eq_12}
\end{equation}
Based on \eqref{eq_12}, we further obtain that 
\begin{equation}
{{\bf{w}}_i}{s_i} = \left( {\sum\limits_{k = 1}^K {{\upsilon _k} \cdot {\bf{h}}_k^H} } \right), \forall i \in {\cal K},
\label{eq_13}
\end{equation}
which is a constant for any $i$. This mathematically verifies that the beamforming vector for one symbol is a phase-rotated version of the beamforming vector for another symbol. Then, with each ${\bf w}_i$ obtained, the beamforming matrix ${\bf W}$ can be obtained and further expressed in a matrix form as
\begin{equation}
\begin{aligned}
{\bf{W}} & = \left[ {{{\bf{w}}_1},{{\bf{w}}_2}, \cdots ,{{\bf{w}}_K}} \right] \\
& = \left( {\sum\limits_{k = 1}^K {{\upsilon_k} \cdot {\bf{h}}_k^H} } \right) \cdot \left[ {s_1^H,s_2^H, \cdots ,s_K^H} \right] \\
& = \left[ {{\bf{h}}_1^H,{\bf{h}}_2^H, \cdots ,{\bf{h}}_K^H} \right]{\left[ {{\upsilon_1},{\upsilon_2}, \cdots ,{\upsilon_K}} \right]^T}\left[ {s_1^H,s_2^H, \cdots ,s_K^H} \right] \\
& = {{\bf{H}}^H}{\bf{\Upsilon}}{{\bf{s}}^H}.
\label{eq_14}
\end{aligned}
\end{equation}
We further express \eqref{eq_3} in a compact form as
\begin{equation}
{\bf{HWs}} = diag\left( {\bf \Lambda} \right){\bf{s}},
\label{eq_15}
\end{equation}
where ${\bf{H}} = {\left[ {{\bf{h}}_1^T,{\bf{h}}_2^T, \cdots ,{\bf{h}}_K^T} \right]^T}$ is the channel matrix and ${\bf \Lambda}  = {\left[ {{\lambda _1},{\lambda _2}, \cdots ,{\lambda _K}} \right]^T}$. By substituting \eqref{eq_14} into \eqref{eq_15}, we can further obtain
\begin{equation}
\begin{aligned}
&{\bf{H}}{{\bf{H}}^H}{\bf{\Upsilon}}{{\bf{s}}^H}{\bf{s}} = diag\left( {\bf \Lambda} \right){\bf{s}} \\
\Rightarrow & {\bf{\Upsilon}} = \frac{1}{K} \cdot {\left( {{\bf{H}}{{\bf{H}}^H}} \right)^{ - 1}}diag\left( {\bf \Lambda} \right){\bf{s}}.
\label{eq_16}
\end{aligned}
\end{equation}
With \eqref{eq_16}, we can obtain the structure of the optimal beamforming matrix as a function of scaling vector $\bf \Lambda$ as
\begin{equation}
{\bf{W}} = \frac{1}{K} \cdot {{\bf{H}}^H}{\left( {{\bf{H}}{{\bf{H}}^H}} \right)^{ - 1}}diag\left( {\bf \Lambda} \right){\bf{s}}{{\bf{s}}^H}.
\label{eq_17}
\end{equation}
It is easy to observe from \eqref{eq_17} that the CI beamforming is a symbol-level beamforming scheme since the beamforming matrix includes the expression of the symbol vector $\bf s$. Moreover, with \eqref{eq_17} the original optimization problem on $\bf W$ is transformed into an optimization on the real-valued scaling vector $\bf \Lambda$. With the fact that $\mu_0 >0$, based on \eqref{eq_9e} we can obtain that the power constraint is strictly active, which leads to 
\begin{equation}
\begin{aligned}
&\left\| {{\bf{Ws}}} \right\|_F^2 = {p_0} \\
\Rightarrow & {\kern 2pt} tr\left\{ {{\bf{Ws}}{{\bf{s}}^H}{{\bf{W}}^H}} \right\} = {p_0} \\
\Rightarrow & {\kern 2pt} {{\bf{s}}^H}{{\bf{W}}^H}{\bf{Ws}} = {p_0}\\
\Rightarrow & {\kern 2pt} \frac{1}{{{K^2}}} \cdot {{\bf{s}}^H}{\bf{s}}{{\bf{s}}^H}diag\left( {\bf \Lambda} \right){\left( {{\bf{H}}{{\bf{H}}^H}} \right)^{ - 1}}diag\left( {\bf \Lambda} \right){\bf{s}}{{\bf{s}}^H}{\bf{s}} = {p_0} \\
\Rightarrow & {\kern 2pt} {{\bf{s}}^H}diag\left( {\bf \Lambda} \right){\left( {{\bf{H}}{{\bf{H}}^H}} \right)^{ - 1}}diag\left( {\bf \Lambda} \right){\bf{s}} = {p_0} \\
\Rightarrow & {\kern 2pt} {{\bf \Lambda}^T}diag\left( {{{\bf{s}}^H}} \right){\left( {{\bf{H}}{{\bf{H}}^H}} \right)^{ - 1}}diag\left( {\bf{s}} \right){\bf \Lambda}  = {p_0} \\
\Rightarrow & {\kern 2pt} {{\bf \Lambda}^T}{\bf T}{\bf \Lambda}  = {p_0},
\label{eq_18}
\end{aligned}
\end{equation}
where we note that $\lambda_k^H = \lambda_k$ as each $\lambda_k$ is real, and $\bf T$ is defined as
\begin{equation}
{\bf T} = diag\left( {{{\bf{s}}^H}} \right){\left( {{\bf{H}}{{\bf{H}}^H}} \right)^{ - 1}}diag\left( {\bf{s}} \right).
\label{eq_19}
\end{equation}
It is easy to obtain that $\bf T$ is Hermitian and positive semi-definite, which further leads to
\begin{equation}
{{\bf \Lambda}^T}{\bf{T}}{\bf \Lambda}  = {{\bf \Lambda}^T}\Re \left( {\bf{T}} \right){\bf \Lambda}  = {{\bf \Lambda}^T}{\bf{V}}{\bf \Lambda}=p_0,
\label{eq_20}
\end{equation}
where ${\bf V} = \Re \left( {\bf{T}} \right)$ is a symmetric and positive semi-definite matrix. With \eqref{eq_20} obtained, we can formulate a new convex optimization problem on ${\bf \Lambda}$ that is equivalent to the original optimization ${\cal P}_1$, expressed as
\begin{equation}
\begin{aligned}
&\mathcal{P}_3: {\kern 3pt} \mathop {\min }\limits_{{{\bf \Lambda}}, {\kern 1pt} t} {\kern 3pt} -t \\
&{\kern 0pt} s. t. {\kern 10pt} {{\bf \Lambda}^T}{\bf{V}}{\bf \Lambda}  - {p_0} = 0 \\
&{\kern 25pt} t - {\lambda _k} \le 0, {\kern 3pt} \forall k \in {\cal K}
\label{eq_21}
\end{aligned}
\end{equation}
The optimal beamforming matrix for the original problem ${\cal P}_1$ in \eqref{eq_5} can be obtained with \eqref{eq_17} based on the obtained $\bf \Lambda$ by solving ${\cal P}_3$. In the following, we analyze the convex optimization ${\cal P}_3$ with the Lagrangian approach, where the Lagrangian of ${\cal P}_3$ is formulated as
\begin{equation}
\begin{aligned}
&{\cal L}\left( {{\bf \Lambda}, t ,{\alpha _0},{\mu _k}} \right) =  - t + {\alpha _0}\left( {{{\bf \Lambda} ^T}{\bf{V}}{\bf \Lambda} - {p_0}} \right) + \sum\limits_{k = 1}^K {{\mu _k}\left( {t - {\lambda _k}} \right)} \\
&=\left( { {\bf 1}^T{\bf u} - 1} \right)t + {\alpha _0} \cdot {{\bf \Lambda}^T}{\bf{V}}{\bf \Lambda}  - {{\bf{u}}^T}{\bf \Lambda}  - {\alpha _0}{p_0},
\label{eq_22}
\end{aligned}
\end{equation} 
where $\alpha_0$ and $\mu_k$ are the dual variables and $\mu_k \ge 0$, $\forall k \in {\cal K}$. ${\bf{u}} = {\left[ {{\mu _1},{\mu _2}, \cdots ,{\mu _K}} \right]^T}$ is a column vector that consists of the dual variables and the vector ${\bf{1}} = {\left[ {1,1, \cdots ,1} \right]^T}$. Based on \eqref{eq_22}, the KKT conditions of ${\cal P}_3$ for optimality are expressed as
\begin{IEEEeqnarray}{rCl} 
\IEEEyesnumber
\frac{{\partial {\cal L}}}{{\partial t}} = {\bf 1}^T{\bf u}  -1 = 0 \IEEEyessubnumber* \label{eq_23a} \\
\frac{{\partial {\cal L}}}{{\partial {\bf \Lambda} }} = {\alpha _0}\left( {{\bf{V}} + {{\bf{V}}^T}} \right){\bf \Lambda}  - {\bf{u}} = {\bf{0}} \label{eq_23b}  \\
\alpha_0 \left({{{\bf \Lambda}^T}{\bf{V}}{\bf \Lambda}  - {p_0}}\right) = 0 \label{eq_23c} \\
{\mu _k}\left( {t - {\lambda _k}} \right) = 0, {\kern 2pt} \forall k \in {\cal K} \label{eq_23d}
\end{IEEEeqnarray}
Based on \eqref{eq_23b}, firstly we have $\alpha_0 \ne 0$, and we can further obtain the expression of $\bf \Lambda$, given by
\begin{equation}
{\bf \Lambda}  = \frac{1}{2\alpha_0}{{\bf{V}}^{ - 1}}{\bf{u}},
\label{eq_24}
\end{equation}
where we note that $\bf V$ is symmetric. With $\alpha_0 \ne 0$, based on \eqref{eq_23c} it is obtained that the power constraint is strictly active, and by substituting \eqref{eq_24} into \eqref{eq_23c}, we can express $\alpha_0$ as a function of the dual vector $\bf u$, given by
\begin{equation}
\begin{aligned}
&{\left( {\frac{1}{{2{\alpha _0}}}{{\bf{V}}^{ - 1}}{\bf{u}}} \right)^T}{\bf{V}}\left( {\frac{1}{{2{\alpha _0}}}{{\bf{V}}^{ - 1}}{\bf{u}}} \right) = {p_0} \\
\Rightarrow &{\kern 2pt} \frac{1}{{4\alpha _0^2}}{{\bf{u}}^T}{{\bf{V}}^{ - 1}}{\bf V}{{\bf{V}}^{ - 1}}{\bf{u}} = {p_0} \\
\Rightarrow &{\kern 2pt} {\alpha _0} = \sqrt {\frac{{{{\bf{u}}^T}{{\bf{V}}^{ - 1}}{\bf{u}}}}{{4{p_0}}}}. 
\label{eq_25}
\end{aligned}
\end{equation}
For the convex optimization ${\cal P}_3$ in \eqref{eq_21}, it is easy to verify that the Slater's condition is satisfied \cite{r27}, which means that the dual gap is zero. Therefore, we can solve ${\cal P}_3$ by solving its corresponding dual problem, which is given by
\begin{equation}
{\cal U} = \mathop {\max }\limits_{{\bf{u}},{\kern 1pt}{\alpha _0}} \mathop {\min }\limits_{{\bf \Lambda}, {\kern 1pt} t} {\cal L}\left( {{\bf \Lambda} ,t,{\alpha _0},{\bf{u}}} \right).
\label{eq_26}
\end{equation}
For the dual problem $\cal U$, the inner minimization is achieved with \eqref{eq_23a} and the obtained ${\bf \Lambda}$ in \eqref{eq_24}, and therefore the dual problem can be further transformed into
\begin{equation}
\begin{aligned}
{\cal U} &=  \mathop {\max }\limits_{{\bf{u}},{\kern 1pt}{\alpha _0}} {\kern 2pt} {\alpha _0}{\left( {\frac{1}{{2{\alpha _0}}}{{\bf{V}}^{ - 1}}{\bf{u}}} \right)^T}{\bf{V}}\left( {\frac{1}{{2{\alpha _0}}}{{\bf{V}}^{ - 1}}{\bf{u}}} \right) \\
& {\kern 12pt} - {{\bf{u}}^T}\left( {\frac{1}{{2{\alpha _0}}}{{\bf{V}}^{ - 1}}{\bf{u}}} \right) - {\alpha _0}{p_0} \\
& =  \mathop {\max }\limits_{{\bf{u}},{\kern 1pt}{\alpha _0}} {\kern 2pt} \frac{1}{{4{\alpha _0}}}{{\bf{u}}^T}{{\bf{V}}^{ - 1}}{\bf V}{{\bf{V}}^{ - 1}}{\bf{u}} - \frac{1}{{2{\alpha _0}}}{{\bf{u}}^T}{{\bf{V}}^{ - 1}}{\bf{u}} - {\alpha _0}{p_0} \\
& = \mathop {\max }\limits_{{\bf{u}},{\kern 1pt}{\alpha _0}} {\kern 1pt} - \frac{1}{{4{\alpha _0}}}{{\bf{u}}^T}{{\bf{V}}^{ - 1}}{\bf{u}} - {\alpha _0}{p_0} \\
& = \mathop {\max }\limits_{{\bf{u}}} {\kern 1pt}  - \frac{{{{\bf{u}}^T}{{\bf{V}}^{ - 1}}{\bf{u}}}}{4{\sqrt {\frac{{{{\bf{u}}^T}{{\bf{V}}^{ - 1}}{\bf{u}}}}{{4{p_0}}}} }} - \sqrt {\frac{{{{\bf{u}}^T}{{\bf{V}}^{ - 1}}{\bf{u}}}}{{4{p_0}}}}  \cdot {p_0} \\
& = \mathop {\max }\limits_{{\bf{u}}} {\kern 1pt} - \sqrt {{p_0} \cdot {{\bf{u}}^T}{{\bf{V}}^{ - 1}}{\bf{u}}}.
\label{eq_27}
\end{aligned}
\end{equation}
Due to the fact that $y = \sqrt x$ is a monotonic function, therefore the dual problem $\cal U$ is equivalent to the following optimization problem
\begin{equation}
\begin{aligned}
&\mathcal{P}_4: {\kern 3pt} \mathop {\min }\limits_{{{\bf u}}} {\kern 3pt} {{\bf{u}}^T}{{\bf{V}}^{ - 1}}{\bf{u}} \\
&{\kern 0pt} s. t. {\kern 10pt} {{\bf{1}}^T}{\bf{u}} = 1 \\
&{\kern 25pt} \mu_k \ge 0, {\kern 3pt} \forall k \in {\cal K}
\label{eq_28}
\end{aligned}
\end{equation}
where the first constraint comes from \eqref{eq_23a}. 

Based on our analysis and transformations above, we have transformed and simplified the original problem, and shown that the original optimization can be solved by solving ${\cal P}_4$. To be more specific, through \eqref{eq_25}, \eqref{eq_24} and \eqref{eq_17}, we arrive at a final closed-form expression of the optimal beamforming matrix as a function of $\bf u$, given by
\begin{equation}
{\bf{W}} = \frac{1}{K} {{\bf{H}}^H}{\left( {{\bf{H}}{{\bf{H}}^H}} \right)^{ - 1}}diag\left\{ {\sqrt {\frac{{{p_0}}}{{{{\bf{u}}^T}{{\bf{V}}^{ - 1}}{\bf{u}}}}} {{\bf{V}}^{ - 1}}{\bf{u}}} \right\}{\bf{s}}{{\bf{s}}^H}.
\end{equation}
Moreover, it is observed that ${\cal P}_4$ is a typical QP optimization problem over a simplex, which can be more efficiently solved with the simplex method \cite{r28} or interior-point methods \cite{r29}, compared to the original CI beamforming formulation ${\cal P}_1$ which is a SOCP optimization.

\subsection{Non-Strict Phase Rotation}
We extend our analysis to the case of non-strict phase rotation. Similarly, before formulating the optimization problem, we firstly construct the condition that the beamforming designs should satisfy such that the received signals are located in the constructive region. Based on Fig. 3, for consistency we denote $\mathop {OA}\limits^ \to   = t \cdot {s_k}$ and $t = |\mathop {OA}\limits^ \to  |$ is the objective to be maximized. Following \eqref{eq_2}, we denote the received signal for user $k$ as $\mathop {OB}\limits^ \to$, which is expressed as
\begin{equation}
\mathop {OB}\limits^ \to   = {{\bf{h}}_k}{\bf{Ws}} = {\lambda _k}{s_k}.
\label{eq_29}
\end{equation}
In the case of non-strict phase rotation, each $\lambda_k$ can be a complex value, which mathematically represents that a phase rotation is included for the received signal $\mathop {OB}\limits^ \to$ compared to the data symbol $s_k$, as shown in Fig. 3. This is different from the case of strict phase rotation where each $\lambda_k$ is strictly real. Then, based on the fact that $\mathop {OC}\limits^ \to$ and $\mathop {CB}\limits^ \to$ are perpendicular, we can obtain the expression of $\mathop {OC}\limits^ \to$ and $\mathop {CB}\limits^ \to$, given by
\begin{equation}
\mathop {OC}\limits^ \to   = \Re \left( {{\lambda _k}} \right){s_k} = \lambda_k^{\Re}{s_k}, {\kern 3pt} \mathop {CB}\limits^ \to   = j \cdot \Im \left( {{\lambda _k}} \right){s_k} = j \cdot \lambda_k^{\Im}{s_k},
\label{eq_30}
\end{equation}
where based on Fig. 3 the imaginary unit `$j$' denotes a phase rotation of $90^{{\rm{o}}}$ geometrically. For simplicity of denotation, we denote $\lambda _k^\Re  = \Re \left( {{\lambda _k}} \right)$ and $\lambda _k^\Im  = \Im \left( {{\lambda _k}} \right)$, respectively. Due to the fact that the nodes `O', `A' and `C' are co-linear, we can further obtain the expression of $\mathop {AC}\limits^ \to$ as
\begin{equation}
\mathop {AC}\limits^ \to   = \left( {\lambda _k^\Re - t} \right){s_k}.
\label{eq_31}
\end{equation}
In Fig. 3, we can observe that to have the received signal $\mathop {OB}\limits^ \to$ located in the constructive region is equivalent to the following condition:
\begin{equation}
\begin{aligned}
&{\theta _{AB}} \le {\theta _t} \\
\Rightarrow & \tan {\theta _{AB}} \le \tan {\theta _t}\\
\Rightarrow & \frac{{|\mathop {CB}\limits^ \to  |}}{{|\mathop {AC}\limits^ \to  |}} = \frac{{\left| {\lambda _k^\Im{s_k}} \right|}}{{\left| {\left( {\lambda _k^\Re - t} \right){s_k}} \right|}} \le \tan {\theta _t} \\
\Rightarrow & \left( {\lambda _k^\Re - t} \right)\tan {\theta _t} \ge \left| {\lambda _k^\Im} \right|.
\label{eq_32}
\end{aligned}
\end{equation}
In the case of $\lambda _k^\Im=0$, $\forall k \in {\cal K}$, \eqref{eq_32} is identical to \eqref{eq_4}, and the non-strict phase rotation reduces to the strict phase rotation. For $\cal M$-PSK modulation, it is observed from the modulation constellation that the threshold angle $\theta_t$ can be expressed as
\begin{equation}
\theta_t = \frac{\pi}{{\cal M}}.
\label{eq_33}
\end{equation}
With the above formulation, we can construct the optimization problem of CI for non-strict phase rotation as
\begin{equation}
\begin{aligned}
&\mathcal{P}_5: {\kern 3pt} \mathop {\max }\limits_{{{\bf{W}}}, {\kern 1pt} t} {\kern 3pt} t \\
&{\kern 0pt} s. t. {\kern 10pt} {{\bf{h}}_k}{\bf Ws} = {\lambda _k}{s_k}, {\kern 3pt} \forall k \in {\cal K} \\
&{\kern 22pt} \left( {\lambda _k^\Re - t} \right)\tan {\theta _t} \ge \left| {\lambda _k^\Im} \right|, {\kern 3pt} \forall k \in {\cal K}\\
&{\kern 22pt} \left\| {{\bf{Ws}}} \right\|_F^2 \le {p_0}
\label{eq_34}
\end{aligned}
\end{equation}
To further analyze the optimization problem for non-strict phase rotation, we first transform ${\cal P}_5$ in \eqref{eq_34} into a standard minimization form, given by
\begin{equation}
\begin{aligned}
&\mathcal{P}_6: {\kern 3pt} \mathop {\min }\limits_{{{\bf{W}}}, {\kern 1pt} t} {\kern 3pt} -t \\
&{\kern 0pt} s. t. {\kern 10pt} {{\bf{h}}_k}{\bf Ws} - {\lambda _k}{s_k}=0, {\kern 3pt} \forall k \in {\cal K} \\
&{\kern 22pt} \left| {\lambda _k^\Im} \right| - \left( {\lambda _k^\Re - t} \right)\tan {\theta _t} \le 0, {\kern 3pt} \forall k \in {\cal K}\\
&{\kern 22pt} \sum\limits_{i = 1}^K {s_i^H{\bf{w}}_i^H{{\bf{w}}_i}{s_i}}  - \frac{p_0}{K} \le 0
\label{eq_35}
\end{aligned}
\end{equation}
Then, by following a similar step in \eqref{eq_8}-\eqref{eq_16} with the Lagrangian approach, we can obtain that the optimal beamforming structure for non-strict phase rotation is the same as that for strict phase rotation, which is given in \eqref{eq_17}. With the power constraint strictly active, we can further obtain that
\begin{equation}
\begin{aligned}
&\left\| {{\bf{Ws}}} \right\|_F^2 = {p_0} \\
\Rightarrow & {\kern 2pt} {{\bf{s}}^H}{{\bf{W}}^H}{\bf{Ws}} = {p_0}\\
\Rightarrow & {\kern 2pt} {{\bf{s}}^H}diag\left( {{{\bf \Lambda}^H}} \right){\left( {{\bf{H}}{{\bf{H}}^H}} \right)^{ - 1}}diag\left( {\bf \Lambda} \right){\bf{s}} = {p_0} \\
\Rightarrow & {\kern 2pt} {{\bf \Lambda}^H}diag\left( {{{\bf{s}}^H}} \right){\left( {{\bf{H}}{{\bf{H}}^H}} \right)^{ - 1}}diag\left( {\bf{s}} \right){\bf \Lambda}  = {p_0} \\
\Rightarrow & {\kern 2pt} {{\bf \Lambda}^H}{\bf T}{\bf \Lambda}  = {p_0},
\label{eq_36}
\end{aligned}
\end{equation}
where $\bf T$ is given by \eqref{eq_19}. However, we note that, different from the case of strict phase rotation, for the case of non-strict phase rotation \eqref{eq_36} is not in a quadratic form since each $\lambda_k$ can be complex. By decomposing
\begin{equation}
{\bf \hat \Lambda}  = {\left[ {\Re \left( {{{\bf \Lambda}^T}} \right),\Im \left( {{{\bf \Lambda}^T}} \right)} \right]^T}, {\kern 3pt} {\bf{\hat T}} = \left[ {\begin{array}{*{20}{c}}
{\Re \left( {\bf{T}} \right)}&{ - \Im \left( {\bf{T}} \right)}\\
{\Im \left( {\bf{T}} \right)}&{\Re \left( {\bf{T}} \right)}
\end{array}} \right],
\label{eq_37}
\end{equation}
we can expand \eqref{eq_36} with its real and imaginary components and further transform the power constraint into a quadratic form, given by
\begin{equation}
\begin{aligned}
&\left\| {{\bf{Ws}}} \right\|_F^2 = {p_0} \\
\Rightarrow & {\kern 2pt} {{\bf \hat \Lambda}^T}{\bf{\hat T}}{\bf \hat \Lambda} - {p_0} = 0.
\label{eq_38}
\end{aligned}
\end{equation}
Similar to the optimization ${\cal P}_3$ in \eqref{eq_21} for strict phase rotation, we can formulate an optimization problem on ${\bf \hat \Lambda}$ for non-strict phase rotation, expressed as
\begin{equation}
\begin{aligned}
&\mathcal{P}_7: {\kern 3pt} \mathop {\min }\limits_{{{\bf \hat \Lambda}}, {\kern 1pt} t} {\kern 3pt} -t \\
&{\kern 0pt} s. t. {\kern 10pt} {{\bf \hat \Lambda}^T}{\bf \hat {T}}{\bf \hat \Lambda}  - {p_0} = 0 \\
&{\kern 24pt} \frac{{\lambda _k^\Im }}{{\tan {\theta _t}}} +t - \lambda _k^\Re \le 0, {\kern 3pt} \forall k \in {\cal K}\\
&{\kern 21pt} - \frac{{\lambda _k^\Im }}{{\tan {\theta _t}}} +t - \lambda _k^\Re \le 0, {\kern 3pt} \forall k \in {\cal K}
\label{eq_39}
\end{aligned}
\end{equation}
where we have transformed the CI constraint with the absolute value on $\lambda_k^{\Im}$ into two separate constraints. We then analyze ${\cal P}_7$ with Lagrangian and KKT conditions, where the Lagrangian of ${\cal P}_7$ is constructed as
\begin{equation}
\begin{aligned}
&{\cal L}\left( {{\bf \hat \Lambda} ,t,{{\hat \alpha }_0},{{\hat \mu }_k},{{\hat \nu }_k}} \right) =  - t + {\hat \alpha _0}\left( {{{\bf \hat \Lambda }^T}{\bf \hat{T}}{\bf \hat \Lambda} - {p_0}} \right) \\
& {\kern 10pt} + \sum\limits_{k = 1}^K {{{\hat \mu }_k}\left( {\frac{{\lambda _k^\Im }}{{\tan {\theta _t}}}  + t - \lambda _k^\Re } \right)}  + \sum\limits_{k = 1}^K {{{\hat \nu }_k}\left( { - \frac{{\lambda _k^\Im }}{{\tan {\theta _t}}}  + t - \lambda _k^\Re } \right)} \\
&= \left[ {\sum\limits_{k = 1}^K {\left( {{{\hat \mu }_k} + {{\hat \nu }_k}} \right)}  - 1} \right]t + {\hat \alpha _0}{{\bf \hat \Lambda}^T}{\bf{\hat T}}{\bf \hat \Lambda} - {\hat \alpha _0}{p_0} \\
& {\kern 10pt} - \sum\limits_{k = 1}^K {\left( {{{\hat \mu }_k} + {{\hat \nu }_k}} \right)\lambda _k^\Re }  + \sum\limits_{k = 1}^K {\left( {{{\hat \mu }_k} - {{\hat \nu }_k}} \right)\frac{{\lambda _k^\Im }}{{\tan {\theta _t}}} },
\label{eq_40}
\end{aligned}
\end{equation} 
where $\hat \alpha_0$, $\hat \mu_k$ and $\hat \nu_k$ are the dual variables, and $\hat \mu_k \ge 0$, $\hat \nu_k \ge 0$, $\forall k$. By introducing
\begin{equation}
\begin{aligned}
&{\bf{\hat u}} = {\left[ {{{\hat \mu }_1},{{\hat \mu }_2}, \cdots ,{{\hat \mu }_K},{{\hat \nu }_1},{{\hat \nu }_2}, \cdots ,{{\hat \nu }_K}} \right]^T}, \\
&{\bf{S}} = \left[ {\begin{array}{*{20}{c}} {  {\bf{I}}}&{-\frac{1}{{\tan {\theta _t}}} \cdot \bf{I}}\\ {  {\bf{I}}}&{  {\frac{1}{{\tan {\theta _t}}} \cdot \bf{I}}}
\end{array}} \right],
\label{eq_41}
\end{aligned}
\end{equation}
where ${\bf \hat u} \in {\cal C}^{2K \times 1}$ and ${\bf S} \in {\cal C}^{2K \times 2K}$, the Lagrangian for ${\cal P}_7$ can be further simplified into
\begin{equation}
{\cal L}\left( {{\bf \hat \Lambda},t,{{\hat \alpha }_0},{\bf{\hat u}}} \right) = \left( {{{\bf{1}}^T}{\bf{\hat u}} - 1} \right)t + {\hat \alpha _0}{{\bf \hat \Lambda}^T}{\bf{\hat T}}{\bf \hat \Lambda} - {{\bf{\hat u}}^T}{\bf{S}}{\bf \hat \Lambda} - {\hat \alpha _0}{p_0}.
\label{eq_42}
\end{equation}
Based on \eqref{eq_42}, we express the KKT conditions for optimality of ${\cal P}_7$ in the following:
\begin{IEEEeqnarray}{rCl} 
\IEEEyesnumber
\frac{{\partial {\cal L}}}{{\partial t}} = {{\bf{1}}^T}{\bf{\hat u}} - 1 = 0  \IEEEyessubnumber* \label{eq_43a} \\
\frac{{\partial {\cal L}}}{{\partial {\bf \hat \Lambda} }} = 2{\hat \alpha _0}{\bf{\hat T}}{\bf \hat \Lambda} - {{\bf{S}}^T}{\bf{\hat u}} = {\bf{0}} \label{eq_43b} \\
\hat \alpha_0 \left({{{\bf \hat \Lambda}^T}{\bf \hat {T}}{\bf \hat \Lambda}  - {p_0}}\right) = 0 \label{eq_43c} \\
{\hat \mu _k}\left( {\frac{{\lambda _k^\Im }}{{\tan {\theta _t}}}  + t - \lambda _k^\Re } \right) = 0, {\kern 3pt} \forall k \in {\cal K} \label{eq_43d} \\
{\hat \nu _k}\left( { - \frac{{\lambda _k^\Im }}{{\tan {\theta _t}}}  + t - \lambda _k^\Re } \right) = 0, {\kern 3pt} \forall k \in {\cal K} \label{eq_43e}
\end{IEEEeqnarray}
Based on \eqref{eq_43b} we can obtain $\hat \alpha_0 \ne 0$ and the expression of $\bf \hat \Lambda$, given by
\begin{equation}
{\bf \hat \Lambda}  =  \frac{1}{{2{{\hat \alpha }_0}}}{{\bf{\hat T}}^{ - 1}}{{\bf{S}}^T}{\bf{\hat u}},
\label{eq_44}
\end{equation}
where we note that $\bf \hat T$ is symmetric. Moreover, from \eqref{eq_43c} we obtain that the power constraint is strictly active with $\hat \alpha_0 \ne 0$, and we can further obtain the expression of $\hat \alpha_0$ as
\begin{equation}
\begin{aligned}
&{\left( { \frac{1}{{2{{\hat \alpha }_0}}}{\bf{\hat T}}{{\bf{S}}^T}{\bf{\hat u}}} \right)^T}{\bf{\hat T}}\left( { \frac{1}{{2{{\hat \alpha }_0}}}{\bf{\hat T}}{{\bf{S}}^T}{\bf{\hat u}}} \right) = {p_0} \\
\Rightarrow & {\kern 2pt} {\hat \alpha _0} = \sqrt {\frac{{{{{\bf{\hat u}}}^T}{\bf{S}}{{{\bf{\hat T}}}^{ - 1}}{{\bf{S}}^T}{\bf{\hat u}}}}{{4{p_0}}}} = \sqrt {\frac{{{{{\bf{\hat u}}}^T}{{\bf{\hat V}}^{ - 1}}{\bf \hat u}}}{{4{p_0}}}}
\label{eq_45}
\end{aligned}
\end{equation}
where for simplicity and consistency we introduce 
\begin{equation}
{{\bf{\hat V}}^{ - 1}} = {\bf{S}}{{\bf{\hat T}}^{ - 1}}{{\bf{S}}^T}.
\label{eq_46}
\end{equation}
Similar to the case for strict phase rotation, it is easy to observe that the Slater's condition is satisfied for ${\cal P}_7$, and therefore by following a similar approach in \eqref{eq_26} and \eqref{eq_27}, the dual problem for ${\cal P}_7$ can be formulated into
\begin{equation}
{\hat {\cal U}} = \mathop {\max }\limits_{{\bf{\hat u}}} {\kern 2pt}  - \sqrt {{p_0} \cdot {{{\bf{\hat u}}}^T}{{{\bf{\hat V}}}^{ - 1}}{\bf{\hat u}}},
\label{eq_47}
\end{equation}
which further leads to the following equivalent optimization for non-strict phase rotation
\begin{equation}
\begin{aligned}
&\mathcal{P}_8: {\kern 3pt} \mathop {\min }\limits_{{{\bf \hat u}}} {\kern 3pt} {{\bf \hat{u}}^T}{{\bf \hat {V}}^{ - 1}}{\bf \hat{u}} \\
&{\kern 0pt} s. t. {\kern 10pt} {{\bf{1}}^T}{\bf \hat{u}} = 1 \\
&{\kern 25pt} \hat u_k \ge 0, {\kern 3pt} \forall k \in \left\{ {1,2, \cdots ,2K} \right\}
\label{eq_48}
\end{aligned}
\end{equation}
where we denote $\hat u_k$ as the $k$-th entry in $\bf \hat u$, and we obtain ${\bf \hat V}^{-1} \in {\cal C}^{2K \times 2K}$ based on \eqref{eq_46}. ${\cal P}_8$ is also a QP optimization over a simplex, which can be efficiently solved. The final optimal beamforming matrix for non-strict phase rotation can be similarly obtained in a closed form as a function of $\bf \hat u$, given by
\begin{equation}
{\bf{W}} = \frac{1}{K} {{\bf{H}}^H}{\left( {{\bf{H}}{{\bf{H}}^H}} \right)^{ - 1}}diag\left\{ { \sqrt {\frac{{{p_0}}}{{{{{\bf{\hat u}}}^T}{{{\bf{\hat V}}}^{ - 1}}{\bf{\hat u}}}}} {\bf{U}}{{{\bf{\hat T}}}^{ - 1}}{{\bf{S}}^T}{\bf{\hat u}}} \right\}{\bf{s}}{{\bf{s}}^H},
\end{equation}
where ${\bf U}=\left[ {\begin{array}{*{20}{c}} {\bf{I}}&{j \cdot {\bf{I}}}\end{array}} \right]$ is a transformation matrix that transform the real-valued vector $\bf \hat \Lambda$ into its complex equivalence.

Based on the formulated equivalent optimization problems ${\cal P}_4$ in \eqref{eq_28} and ${\cal P}_8$ in \eqref{eq_48}, we note the similarity of the optimization problem for strict phase rotation and non-strict phase rotation. We observe that the objective function of ${\cal P}_4$ for strict phase rotation and ${\cal P}_8$ for non-strict phase rotation is identical, and both optimization problems share the same constraints. It is further observed that the only difference between ${\cal P}_4$ and ${\cal P}_8$ is the problem size. It is then concluded that a $K$-dimensional optimization problem for non-strict phase rotation and a $2K$-dimensional optimization for strict phase rotation share the same problem formulation, and therefore they can be solved in a similar way. 

\section{CI as a Generalization of ZF Precoding}
In this section, we discuss the connection between the CI beamforming for strict phase rotation and the conventional ZF precoding. For the CI beamforming with non-strict phase rotation, the connection can be obtained in a similar way. To compare the CI beamforming and the conventional ZF precoding, as a reference we first present the precoded signal vector of ZF, given by
\begin{equation}
{\bf x}_{ZF}={{\bf{W}}_{ZF}} {\bf s}= \frac{1}{f} \cdot {{\bf{H}}^H}{\left( {{\bf{H}}{{\bf{H}}^H}} \right)^{ - 1}}{\bf{s}},
\label{eq_49}
\end{equation}  
where $f$ is the scaling factor to meet the transmit power constraint. For fairness of comparison, we employ a symbol-level normalization for ${\bf W}_{ZF}$ such that $\left\| {{{\bf{W}}_{ZF}}{\bf{s}}} \right\|_F^2 = {p_0}$ as for the considered CI beamforming, which leads to the expression of $f$ as
\begin{equation}
f = \sqrt {\frac{{\left\| {{{\bf{W}}_{ZF}}{\bf{s}}} \right\|_F^2}}{{{p_0}}}}  = \sqrt {\frac{{{{\bf{s}}^H}{\left( {{\bf{H}}{{\bf{H}}^H}} \right)^{ - 1}}{\bf{s}}}}{{{p_0}}}}.
\label{eq_50}
\end{equation}
By denoting ${\bf C}=\left( {{\bf HH}^{H}}\right)^{-1}$, the expression of $f$ can be further transformed into
\begin{equation}
\begin{aligned}
&f = \sqrt {\frac{{\sum\limits_{m = 1}^K {\sum\limits_{n = 1}^K {{\bf{C}}\left( {m,n} \right)s_m^H{s_n}} } }}{{{p_0}}}} \\
\Rightarrow & {{\sum\limits_{m = 1}^K {\sum\limits_{n = 1}^K {{\bf{C}}\left( {m,n} \right)s_m^H{s_n}} } }} = f^2 p_0.
\label{eq_51}
\end{aligned}
\end{equation}
Subsequently, we perform the mathematical analysis on the optimization problem ${\cal P}_4$ on $\bf u$ for strict phase rotation. By applying the Lagrangian approach, we can obtain the Lagrangian of ${\cal P}_4$, given by
\begin{equation}
\begin{aligned}
{\cal L} \left( {{\bf{u}},{q_0},{\bf{q}}} \right) &= {{\bf{u}}^T}{{\bf{V}}^{ - 1}}{\bf{u}} + {q_0}\left( {{{\bf{1}}^T}{\bf{u}} - 1} \right) - \sum\limits_{k = 1}^K {{q_k}{\mu _k}} \\
&= {{\bf{u}}^T}{{\bf{V}}^{ - 1}}{\bf{u}} + q_0 \cdot {\bf 1}^T{\bf u} -{\bf q}^T {\bf u} -q_0,
\label{eq_52}
\end{aligned}
\end{equation}
where the vector ${\bf{q}} = {[{q_1},{q_2}, \cdots ,{q_K}]^T}$ consists of each non-negative dual variable $q_k$ of ${\cal P}_4$. Based on \eqref{eq_52}, we express the KKT conditions of ${\cal P}_4$ as
\begin{IEEEeqnarray}{rCl} 
\IEEEyesnumber
\frac{{\partial {\cal L}}}{{\partial {\bf{u}}}} = 2{{\bf{V}}^{ - 1}}{\bf{u}} + {q_0} \cdot {\bf{1}} - {\bf{q}} = {\bf{0}}  \IEEEyessubnumber* \label{eq_53a} \\
{q_0}\left( {{{\bf{1}}^T}{\bf{u}} - 1} \right) = 0 \label{eq_53b} \\
{q_k}{\mu _k} = 0, {\kern 3pt} \forall k \in {\cal K} \label{eq_53c}
\end{IEEEeqnarray}
Based on \eqref{eq_53a} we can obtain the expression of $\bf u$ as a function of the dual variables, given by 
\begin{equation}
{\bf{u}} = \frac{1}{2}{\bf{V}}\left( {{\bf{q}} - {q_0} \cdot {\bf{1}}} \right),
\label{eq_54}
\end{equation}
and each $\mu_k$ as
\begin{equation}
{\mu _k} = \frac{1}{2} \left( {{{\bf{v}}_k}{\bf{q}} - {q_0}{a_k}} \right), \forall k \in {\cal K},
\label{eq_55}
\end{equation}
where we have decomposed $\bf V$ into ${\bf{V}} = {\left[ {{\bf{v}}_1^T,{\bf{v}}_2^T, \cdots ,{\bf{v}}_K^T} \right]^T}$. ${\bf{a}} = {[{a_1},{a_2}, \cdots ,{a_K}]^T}$ denotes the column vector obtained from the sum of $\bf V$ by column, with each entry given by
\begin{equation}
{a}_k = \sum\limits_{i = 1}^K {{\bf{V}}\left( {k,i} \right)}.
\label{eq_56}
\end{equation}
By substituting the expression of $\bf u$ into (53b), we further obtain that
\begin{equation}
\begin{aligned}
& {{\bf{1}}^T}{\bf{u}} - 1 = 0 \\
\Rightarrow & \frac{1}{2}\sum\limits_{k = 1}^K {\left\{ {\left[ {\sum\limits_{i = 1}^K {{\bf{V}}\left( {k,i} \right){q_i}} } \right] - {q_0}{a_k}} \right\}} -1 =0  \\
\Rightarrow & \frac{1}{2}\sum\limits_{i = 1}^K {\left[ {\sum\limits_{k = 1}^K {{\bf{V}}\left( {k,i} \right)} } \right]} {q_i} - \frac{1}{2}{q_0}c - 1 = 0 \\
\Rightarrow & \frac{1}{2}\sum\limits_{i = 1}^K {{b_i}{q_i}}  - \frac{1}{2}{q_0}c - 1 = 0 \\
\Rightarrow & {q_0} = \frac{{{{\bf{b}}}{\bf{q}} - 2}}{c},
\label{eq_57}
\end{aligned}
\end{equation}
where $\bf b$ is a row vector obtained from the sum of $\bf V$ by row and $c$ denotes the sum of all the entries in $\bf V$. $\bf b$ and $c$ are given by
\begin{equation}
\begin{aligned}
&{\bf{b}} = \sum\limits_{k = 1}^K {{{\bf{v}}_k}} = {\bf a}^T, \\
&c  = \sum\limits_{k = 1}^K {\sum\limits_{i = 1}^K {{\bf{V}}\left( {k,i} \right)} } = {\bf a}^T {\bf 1},
\label{eq_58}
\end{aligned}
\end{equation}
where ${\bf b}={\bf a}^T$ is based on the fact that $\bf V$ is symmetric. By substituting the expression of $q_0$ in \eqref{eq_57} into \eqref{eq_55}, the expression of each $\mu_k$ can be further transformed into
\begin{equation}
\begin{aligned}
{\mu _k} &= \frac{1}{2}{{\bf{v}}_k}{\bf{q}} - \frac{{{a_k}}}{2}\frac{{{{\bf a}^T{\bf q}} - 2}}{c} \\
&= \frac{1}{2}\left( {{{\bf{v}}_k} - \frac{{{a_k}}}{c}{\bf{a}}^T} \right){\bf{q}} + \frac{{{a_k}}}{c},
\label{eq_59}
\end{aligned}
\end{equation}
which further leads to the expression of $\bf u$ as
\begin{equation}
{\bf{u}} = \frac{1}{2}\left( {{\bf{V}} - {\bf \Phi} } \right){\bf{q}} + \frac{{\bf{a}}}{c},
\label{eq_60}
\end{equation}
where ${\bf \Phi}  = \frac{{\bf{aa}}^T}{c}$. By substituting the expression of $\bf u$ into the expression of $\bf \Lambda$ in \eqref{eq_24}, we can further obtain that
\begin{equation}
\begin{aligned}
{\bf \Lambda} &= \frac{1}{{2{\alpha _0}}}{{\bf{V}}^{ - 1}}\left[ {\frac{1}{2}\left( {{\bf{V}} - {\bf \Phi}} \right){\bf{q}} + \frac{{\bf{a}}}{c}} \right] \\
&= \frac{1}{{2{\alpha _0}c}} {{\bf{V}}^{ - 1}}{\bf{a}} + \frac{1}{{4{\alpha _0}}}\left( {{\bf{I}} - {{\bf{V}}^{ - 1}}{\bf \Phi} } \right){\bf{q}} \\
&= \frac{1}{{2{\alpha _0}c}} {\bf m} + \frac{1}{{4{\alpha _0}}}\left( {{\bf{I}} - {{\bf{V}}^{ - 1}}{\bf \Phi} } \right){\bf{q}},
\label{eq_61}
\end{aligned}
\end{equation}
where we have defined 
\begin{equation}
{\bf{m}} = {{\bf{V}}^{ - 1}}{\bf{a}}.
\label{eq_62}
\end{equation}
In \eqref{eq_62}, ${\bf m} \in {\cal C}^{K \times 1}$ and ${\bf{m}} = {\left[ {{m_1},{m_2}, \cdots ,{m_K}} \right]^T}$. Based on the expression of $\bf a$, each $m_k$ is obtained as
\begin{equation}
\begin{aligned}
m_k &= \sum\limits_{n = 1}^K {{{\bf{V}}^{ - 1}}\left( {k,n} \right){a_n}} \\
&= \sum\limits_{n = 1}^K {{{\bf{V}}^{ - 1}}\left( {k,n} \right)\sum\limits_{i = 1}^K {{\bf{V}}\left( {n,i} \right)} } \\
&= \sum\limits_{n = 1}^K {{{\bf{V}}^{ - 1}}\left( {k,n} \right){\bf{V}}\left( {n,k} \right)}  + \sum\limits_{i \ne k} {\sum\limits_{n = 1}^K {{{\bf{V}}^{ - 1}}\left( {k,n} \right){\bf{V}}\left( {n,i} \right)} } \\
&= 1 + \sum\limits_{i \ne k} 0 \\
&= 1,
\label{eq_63}
\end{aligned}
\end{equation}
which also means that ${\bf m}={\bf V}^{-1}{\bf a}={\bf 1}$. With this fact, the expression of $\bf \Lambda$ is further transformed into
\begin{equation}
{\bf \Lambda} = \frac{1}{{2{\alpha _0}c}}{\bf{1}} + \frac{1}{{4{\alpha _0}}}\left( {{\bf{I}} - {{\bf{V}}^{ - 1}}{\bf \Phi} } \right){\bf{q}},
\label{eq_64}
\end{equation}
based on which we shall discuss the connection between the CI beamforming and the conventional ZF scheme. In \eqref{eq_64}, if we set 
\begin{equation}
{q_k} = 0, {\kern 3pt} \forall k \in {\cal K},
\label{eq_65}
\end{equation}
based on \eqref{eq_60} we can obtain that 
\begin{equation}
{\bf{u}} = \frac{{\bf{a}}}{c}
\label{eq_66}
\end{equation}
and based on \eqref{eq_25} we further obtain that
\begin{equation}
{\alpha _0} = \sqrt {\frac{{{{\bf{a}}^T}{{\bf{V}}^{ - 1}}{\bf{a}}}}{{4{c^2}{p_0}}}}  = \sqrt {\frac{{{{\bf{a}}^T}{\bf{1}}}}{{4{c^2}{p_0}}}}  = \sqrt {\frac{c}{{4{c^2}{p_0}}}}  = \frac{1}{{2\sqrt {c{p_0}} }}.
\label{eq_67}
\end{equation}
Then, the expression of ${\bf \Lambda}$ is simplified into
\begin{equation}
{\bf \Lambda} = \frac{1}{{2{\alpha _0}c}}{\bf{1}} = \frac{{2\sqrt {c{p_0}} }}{{2c}} = \sqrt {\frac{{{p_0}}}{c}}.
\label{eq_68}
\end{equation}
Based on the expression of $\bf T$ in \eqref{eq_19}, we can obtain the expression of ${\bf{T}}\left( {m,n} \right)$ as
\begin{equation}
{\bf{T}}\left( {m,n} \right) = {\bf{C}}\left( {m,n} \right)s_m^H{s_n},
\label{eq_69}
\end{equation}
and with the fact that $\bf T$ is Hermitian, we further obtain that
\begin{equation}
\begin{aligned}
c &= \sum\limits_{m = 1}^K {\sum\limits_{n = 1}^K {{\bf{V}}\left( {m,n} \right) = } } \sum\limits_{m = 1}^K \sum\limits_{n = 1}^K {\bf{T}}\left( {m,n} \right) \\
&= \sum\limits_{m = 1}^K {\sum\limits_{n = 1}^K {{\bf{C}}\left( {m,n} \right)s_m^H{s_n}} }  \\
&= {f^2}{p_0}.
\label{eq_70}
\end{aligned}
\end{equation}
By substituting \eqref{eq_70} into \eqref{eq_68}, we obtain
\begin{equation}
{\bf \Lambda}=\sqrt {\frac{{{p_0}}}{{{f^2}{p_0}}}}  = \frac{1}{f}.
\label{eq_71}
\end{equation}
In this case, with all dual variables equal to zero, each $\lambda_k$ is real and identical, which further leads to the expression of the precoded signal vector for CI as
\begin{equation}
\begin{aligned}
{{\bf{x}}_{CI}} &= {\bf{Ws}} = \frac{1}{K} \cdot {{\bf{H}}^H}{\left( {{\bf{H}}{{\bf{H}}^H}} \right)^{ - 1}}\frac{1}{f}{\bf{s}}{{\bf{s}}^H}{\bf{s}} \\
&=\frac{1}{f} \cdot {{\bf{H}}^H}{\left( {{\bf{H}}{{\bf{H}}^H}} \right)^{ - 1}}{\bf{s}} \\
&={\bf x}_{ZF},
\label{eq_72}
\end{aligned}
\end{equation}
which is identical to the precoded signal vector based on ZF, where we denote ${\bf x}_{CI}$ as the transmit signal vector for the CI beamforming. 

The above results show that the conventional ZF precoding can be regarded as a special case of the CI beamforming with all the dual variables being zero, as demonstrated in \eqref{eq_65}. The performance of ZF method is therefore the lower-bound of the CI beamforming. We shall discuss under what conditions the CI beamforming is equivalent to the ZF approach in the following section. It can be further observed that the performance of the CI beamforming will be superior to the ZF scheme if not all the dual variables are zero, as shown in \eqref{eq_64} where the existence of non-zero dual variables will increase the minimum value in $\bf \Lambda$. We further note that when the optimality is achieved, the minimum value in $\bf \Lambda$ is guaranteed to be not smaller than \eqref{eq_71}, for otherwise the ZF beamforming will generate a larger minimum value in $\bf \Lambda$, which means that ZF should be the optimal and this causes contradiction.

\section{Proposed Iterative Closed-Form Scheme}
In this section, our proposed iterative close-form scheme is introduced. Throughout this section, we consider the case of strict phase rotation, while the extension to the non-strict phase rotation is trivial and briefly included, as both optimization problems share the same problem formulation, discussed in Section IV. To introduce the proposed scheme, we first transform the expression of $\bf u$ in \eqref{eq_60} into
\begin{equation}
{\bf u}= \frac{1}{2}{\bf G} {\bf{q}} + \frac{{\bf{a}}}{c},
\label{eq_73}
\end{equation}
where $\bf G$ is defined as
\begin{equation}
{\bf G}= {\bf V} - {\bf \Phi}.
\label{eq_74}
\end{equation}
Then, based on the optimality conditions in (55), as long as we find a $\bf u$ and the corresponding dual vector $\bf q$ that satisfy (55), the obtained $\bf u$ is the optimal solution for ${\cal P}_4$. This further leads to the following optimization problem
\begin{equation}
\begin{aligned}
&\mathcal{P}_9: {\kern 3pt} \mathop {{\rm{find}}}\limits_{\bf{q}} {\kern 3pt} {\bf{u}} \\
&{\kern 0pt} s. t. {\kern 8pt} {\bf u}= \frac{1}{2}{\bf G} {\bf{q}} + \frac{{\bf{a}}}{c} \\
&{\kern 22pt} {{\bf{1}}^T}{\bf{u}} - 1 = 0 \\
&{\kern 22pt} {\mu _k}{q_k} = 0, {\kern 2pt} {\mu _k} \ge 0, {\kern 2pt} {q_k} \ge 0, {\kern 2pt} \forall k \in {\cal K}
\label{eq_75}
\end{aligned}
\end{equation}
For clarity of description, we define a set $\cal S$ as
\begin{equation}
{\cal S} = \left\{ {k {\kern 3pt} | {\kern 3pt} {a_k} < 0, {\kern 2pt} \forall k \in {\cal K}} \right\}.
\label{eq_76}
\end{equation}
In the following based on $\cal S$ we discuss the solution of ${\cal P}_9$ and propose the iterative closed-form scheme.

\subsection{${\cal S} = \emptyset$}
When ${\cal S} = \emptyset$, this means $a_k \ge 0$, $\forall k \in {\cal K}$. Then, based on the fact that ${\bf 1}^T{\bf a}=c$ in \eqref{eq_58}, it is obvious that
\begin{equation}
{\bf u} = \frac{\bf a}{c}, {\kern 3pt} q_k=0, {\kern 2pt} \forall k \in {\cal K}
\label{eq_77}
\end{equation}
satisfies all the conditions in \eqref{eq_75}. Therefore, when ${\cal S} = \emptyset$, the optimal ${\bf u}^*$ and ${\bf q}^*$ can be obtained as 
\begin{equation}
{\bf u}^* = \frac{\bf a}{c}, {\kern 2pt} {\bf q}^*= {\bf 0}.
\label{eq_78}
\end{equation} 
In this case, based on our analysis in Section IV the CI beamforming is identical to the ZF approach, where no performance gains can be obtained.

\subsection{$card\left( {\cal S} \right) \ne \emptyset$}
When $card\left( {\cal S} \right) \ne \emptyset$, this means that there is at least one entry in $\bf a$ that is smaller than zero. It is then obvious that the optimal ${{\bf{u}}^*} \ne \frac{{\bf{a}}}{c}$ due to the requirement that $\mu_k \ge 0$. In this case, we can obtain that not all the dual variables are zero, and we need to introduce at least one positive $q_i$ such that each $\mu_k \ge 0$. We firstly set
\begin{equation}
{{\bf{u}}} = \frac{{\bf{a}}}{c}, {\kern 3pt} {\bf q}={\bf 0}.
\label{eq_79}
\end{equation} 
Subsequently, we sort the entries in ${{\bf{u}}}$ following an ascending order, expressed as
\begin{equation}
{\bf{d}} = {\rm{sort}}\left( {{{\bf{u}}}} \right),
\label{eq_80}
\end{equation}
where ${\bf{d}}$ is the sorted vector and ${\bf{d}} = {\left[ {{d_1},{d_2}, \cdots ,{d_K}} \right]^T}$. ${\rm{sort}}\left( {\cdot} \right)$ denotes the sort function, and without loss of generality we denote $k$ as the minimum value in $\bf u$, which leads to
\begin{equation}
{\mu_k} = \min \left( {\bf{u}} \right) = {d}_1.
\label{eq_81}
\end{equation}
With \eqref{eq_81} we can also obtain $a_k < 0$ and ${d}_1 < 0$. Let us firstly introduce only one positive dual variable $q_k$ that corresponds to $\mu_k$ while keeping other dual variables zero. Based on the complementary slackness condition, when $q_k \ne 0$, we obtain $\mu_k=0$, and this further leads to 
\begin{equation}
\begin{aligned}
& {\mu _k} = \frac{1}{2}\sum\limits_{i = 1}^K {{\bf{G}}\left( {k,i} \right)} {{q_i}} + \frac{{{a_k}}}{c} =0 \\
\Rightarrow & \frac{1}{2}{\bf{G}}\left( {k,k} \right){q_k} + \frac{{{a_k}}}{c} = 0 \\
\Rightarrow & {q_k} =  - \frac{{2{a_k}}}{{{\bf{G}}\left( {k,k} \right)c}}.
\label{eq_82}
\end{aligned}
\end{equation}
where based on the definition of $\bf G$ we can verify that ${\bf G}\left( {k,k} \right)>0$. Based on the fact that $a_k <0$ it is then obtained that $q_k >0$. We further define a vector $\bf i$
\begin{equation}
{\bf{i}} = \left[ k \right], {\kern 2pt} \forall q_k \ne 0,
\label{eq_83}
\end{equation}
and a set $\cal I$ that consists of all the entries in $\bf i$, where we denote ${\bf{i}} = \left[ {{i_1},{i_2} \cdots ,{i_M}} \right]$ and $card\left( {\cal I} \right) = M$. By updating ${\bf q}$ with the updated $q_k$ based on \eqref{eq_82}, the updated $\bf u$ can be expressed as
\begin{equation}
{\bf{u}} = \frac{1}{2}\sum\limits_{k \in {\cal I}} {{{\bf{g}}_k}{q_k}}  + \frac{{\bf{a}}}{c},
\label{eq_84}
\end{equation}
where we have decomposed ${\bf{G}} = [{{\bf{g}}_1},{{\bf{g}}_2}, \cdots ,{{\bf{g}}_K}]$, and \eqref{eq_84} satisfies $\mu_k=0$. We verify whether the minimum value in the updated $\bf u$ satisfies the non-negative condition, and the updated $\bf u$ is the optimal solution of ${\cal P}_9$ if $\min \left( {\bf{u}} \right)$ is non-negative. If this condition is not satisfied, this means that one dual variable is not enough and we need to introduce an additional dual variable. In this case, we first sort the updated $\bf u$ based on \eqref{eq_80} and then find the minimum value in the updated $\bf d$, where without loss of generality we denote
\begin{equation}
{\mu_l} = {d_1}, {\kern 3pt} q_l \ne 0,
\label{eq_85}
\end{equation}
where we note that $d_1$ in \eqref{eq_85} is different from $d_1$ in \eqref{eq_81} as $\bf u$ has been updated. With the existence of two non-zero dual variables, we obtain ${\bf{i}} = \left[ {k,l} \right]$ and ${\cal I}=\left\{ {k, {\kern 1pt} l}\right\}$. We can then formulate a matrix ${\bf Z} \in {\cal C}^{card\left( {\cal S} \right) \times card\left( {\cal S} \right)}$ as
\begin{equation}
{\bf Z} = \left[ {\begin{array}{*{20}{c}}
{{\bf{G}}\left( {k,k} \right)}&{{\bf{G}}\left( {k,l} \right)}\\
{{\bf{G}}\left( {l,k} \right)}&{{\bf{G}}\left( {l,l} \right)}
\end{array}} \right].
\label{eq_86}
\end{equation}
By defining 
\begin{equation}
{\bf{\tilde q}} = {\left[ {{q_k},{q_l}} \right]^T}, {\kern 2pt} {\bf{\tilde a}} = {\left[ {{a_k},{a_l}} \right]^T}, {\kern 2pt} {\bf{\tilde u}} = {\left[ {{\mu _k},{\mu _l}} \right]^T}
\label{eq_87}
\end{equation}
that consists of the entries that correspond to the numbers of non-zero dual variables, we obtain 
\begin{equation}
{\bf{\tilde u}} = {\bf 0},
\label{eq_88}
\end{equation}
which is due to the complementary slackness condition. With \eqref{eq_84} and \eqref{eq_88}, we can further obtain ${\bf{\tilde q}}$ as
\begin{equation}
\begin{aligned}
& {\kern 2pt} {\bf{\tilde u}} = \frac{1}{2}\left( { \left[ {\begin{array}{*{20}{c}}
{{\bf{G}}\left( {k,k} \right)}\\
{{\bf{G}}\left( {l,k} \right)}
\end{array}} \right]{q_k} + \left[ {\begin{array}{*{20}{c}}
{{\bf{G}}\left( {k,l} \right)}\\
{{\bf{G}}\left( {l,l} \right)}
\end{array}} \right]{q_l}} \right) + \frac{{\bf{\tilde a}}}{c} = {\bf 0} \\
\Rightarrow & {\kern 2pt} \frac{1}{2}{\bf{Z\tilde q}} + \frac{{{\bf{\tilde a}}}}{c} = {\bf 0} \\
\Rightarrow & {\kern 2pt} {\bf{\tilde q}} =  - \frac{2}{c} \cdot {{\bf{Z}}^{ - 1}}{\bf{\tilde a}}.
\label{eq_89}
\end{aligned}
\end{equation}
If each entry in the obtained ${\bf{\tilde q}}$ satisfies the non-negative condition, we update $\bf u$ based on \eqref{eq_84} with the updated ${\cal I}$, and further check whether the minimum value in the updated $\bf u$ satisfies the non-negative condition.

Nevertheless, when $card \left( {\cal S}\right)>1$, it cannot be guaranteed that each entry in the obtained ${\bf{\tilde q}}$ in \eqref{eq_89} satisfies the non-negative condition. In this case, a retraction approach is required. To be more specific, if there is one entry in the obtained ${\bf{\tilde q}}$ that is negative, we firstly find the corresponding number of the negative dual variable, given by
\begin{equation}
k = {\rm{find}}\left( {{q_k} < 0} \right),
\label{eq_90}
\end{equation}
where the `find' function returns the index of the negative entry in $\bf q$. We further obtain the corresponding column index of $k$ in ${\bf{i}}$, expressed as
\begin{equation}
{i_m} = k.
\label{eq_91}
\end{equation}
We then reset 
\begin{equation}
{\bf{i}} = \left[ {{i_1},{i_2}, \cdots ,{i_{m-1}}} \right], 
\label{eq_92}
\end{equation}
which means that there are currently $(m-1)$ positive dual variables and we set all the obtained dual variables obtained after $(m-1)$ to $0$. With \eqref{eq_92}, we reformulate the corresponding ${\cal I}$ and $\bf u$. Then, for the $m$-th dual variable, instead of selecting the number that corresponds to the minimum value in $\bf d$ as in \eqref{eq_81}, we select $\mu_m$ that corresponds to the second minimum value in $\bf d$. Based on \eqref{eq_80} we obtain
\begin{equation}
\mu_m=d_2,
\label{eq_93}
\end{equation}
and we update ${\bf i}$ and $\cal I$. With the updated $\cal I$, we calculate ${\bf{\tilde q}}$ based on \eqref{eq_89}, and we repeat the above process \eqref{eq_80}-\eqref{eq_93} by increasing the number of non-zero dual variables one at a time until all the entries in the updated $\bf u$ are non-negative, on the condition that in each step the entries in the obtained ${\bf{\tilde q}}$ are non-negative. 

\begin{algorithm}[!b]
  \caption{Proposed Iterative Closed-form Scheme for Strict Phase Rotation}
  \begin{algorithmic}
    \State ${\bf input:}$ ${\bf s}$, $\bf H$
    \State ${\bf output:}$ ${\bf W}^*$
    \State Initialize ${\bf i}=[{\kern 2pt}]$, ${\cal I}=\emptyset$, ${\bf N}=[ 1 ]$, $t=1$, and $n=0$;
    \State Calculate $\bf T$ based on \eqref{eq_19}; Obtain ${\bf V} = \Re \left( {\bf T} \right)$;
    \State Calculate $\bf a$ based on \eqref{eq_56} and $c$ based on \eqref{eq_58};
    \State Calculate ${\bf G}={\bf V}-\frac{{\bf aa}^T}{c}$; Calculate ${\bf u}=\frac{\bf a}{c}$;
    \State Obtain $\cal S$ based on \eqref{eq_76};
    \If {$card\left( {\cal S}\right) = \emptyset$}
    \State Obtain ${\bf u}^*={\bf u}$;
    \Else
    \While {$\min \left( {\bf{u}} \right) < 0$ and $n<n_{\rm max}$}
    \State ${\bf{d}} = {\rm{sort}}\left( {{{\bf{u}}}} \right)$;
    \State find $k$ such that $\mu_k=d_{t}$;
    \State Stack ${\bf{N}} = \left[ {\begin{array}{*{20}{c}}{\bf{N}}&1
\end{array}} \right]$;
    \State Update $\bf i$ and $\cal I$; Formulate ${\bf Z}$ based on $\cal I$ and $\bf G$;
    \State Calculate ${\bf{\tilde q}}$ based on $\cal I$ and ${\bf Z}$ with \eqref{eq_89};
    \If {$\min \left( {{\bf{\tilde q}}} \right) \ge 0$}
    \State Update ${\bf{u}}$ based on \eqref{eq_84};
    \State $t=1$;
    \Else
    \State find $k$ such that ${q_k} = \min \left( {{\bf{\tilde q}}} \right)$;
    \State find $m$ such that $i_m=k$;
    \State Set $\bf i$ based on \eqref{eq_92}; Update $\cal I$;
    \State Formulate ${\bf Z}$ based on $\cal I$ and $\bf G$;
    \State Update ${\bf{\tilde q}}$ with \eqref{eq_89}; Update ${\bf u}$ with \eqref{eq_83};
    \State Reformulate ${\bf{N}} = {\bf{N}}\left( {1:m} \right)$; 
    \State Update ${\bf{N}}\left( {m} \right) \leftarrow {\bf{N}}\left( {m} \right) + 1$;
    \State Update $t = {\bf{N}}\left( {m} \right)$;
    \EndIf
    \State $n \leftarrow n+1$;
    \EndWhile
    \State Obtain ${\bf u}^*={\bf u}$;
    \EndIf
    \State Calculate ${\bf W}^*$ based on the obtained ${\bf u}^*$ with (29).
   \end{algorithmic}
\end{algorithm}

\subsection{The Iterative Algorithm}
Based on the above description, we summarize the proposed scheme for strict phase rotation in Algorithm 1. Since the algorithm will find the $\bf u$ and $\bf q$ that satisfy the KKT conditions for optimality, the obtained $\bf u$ is therefore the optimal solution to the optimization problem ${\cal P}_4$ for strict phase rotation, and the optimal beamforming matrix can be obtained with (29) accordingly. While the algorithm is for the case of the strict phase rotation, it is trivial to extend to the case of non-strict phase rotation by substituting ${\bf V}$ with ${\bf \hat V}$ in \eqref{eq_46} to obtain the optimal ${\bf \hat u}^*$. Subsequently, the optimal beamforming matrix for non-strict phase rotation is obtained with (50).

We note that, while the above algorithm includes an iterative design, within each iteration a closed-form solution is indeed obtained and the algorithm only includes linear matrix manipulations, which is computationally efficient. Moreover, it will be shown that the number of iterations required is small, especially when the number of users is small. We further note that, while the KKT optimality conditions are not satisfied before the final iteration, the solution obtained within each iteration is indeed a feasible solution that satisfies the power constraint for the beamforming and achieves an improved performance over ZF but a sub-optimal performance compared to the optimal CI beamforming. Indeed, the obtained beamformer in each iteration approaches the optimal beamforming strategy with the increasing iteration number. Therefore, the proposed iterative scheme can also achieve a flexible performance-complexity tradeoff by limiting the maximum number of iterations $n_{\rm max}$ in Algorithm 1. The complexity gain of the proposed scheme and the performance-complexity tradeoff will both be numerically studied in the following section.

\section{Numerical Results}
In this section, the numerical results of the proposed schemes are presented and compared with the traditional CI beamforming based on the Monte Carlo simulations. In each plot, we assume the total transmit power available as $p_0=1$, and the transmit SNR per antenna as $\rho = {1 \mathord{\left/ {\vphantom {1 {{\sigma ^2}}}} \right. \kern-\nulldelimiterspace} {{\sigma ^2}}}$. We compare our proposed iterative schemes with the traditional closed-form ZF-based methods, optimization-based SINR balancing approaches \cite{r7}\cite{r10}, and CI beamforming approaches ${\cal P}_1$ and ${\cal P}_5$ for both strict and non-strict phase rotation.

For clarity the following abbreviations are used throughout this section:

\begin{enumerate}
\item `ZF': traditional ZF scheme with symbol-level power normalization in \eqref{eq_49} and \eqref{eq_50};
\item `RZF': traditional RZF scheme with symbol-level power normalization, where the precoded signal vector is given by
\begin{equation}
{{\bf{x}}_{RZF}} = {\bf W}_{RZF}{\bf s} = \frac{1}{f} \cdot {{\bf{H}}^H}{\left( {{\bf{H}}{{\bf{H}}^H} + \frac{K}{\rho } \cdot {\bf{I}}} \right)^{ - 1}}{\bf{s}}
\label{eq_94}
\end{equation}
with the symbol-level scaling factor $f$ given by 
\begin{equation}
f = \frac{{{{\left\| {{{\bf{W}}_{RZF}}{\bf{s}}} \right\|}_F}}}{{\sqrt {{p_0}} }};
\label{eq_95}
\end{equation}
\item `SINR Balancing': the SINR balancing approach based on bisection search method \cite{r7};
\item `CI-opt, Strict/Non-Strict': traditional CI beamforming, ${\cal P}_1$ for strict phase rotation and ${\cal P}_5$ for non-strict phase rotation;
\item `CI-CF, Strict/Non-Strict': the proposed iterative closed-form scheme for strict/non-strict phase rotation based on Algorithm 1.
\end{enumerate}

\begin{figure}[!t]
\centering
\includegraphics[scale=0.45]{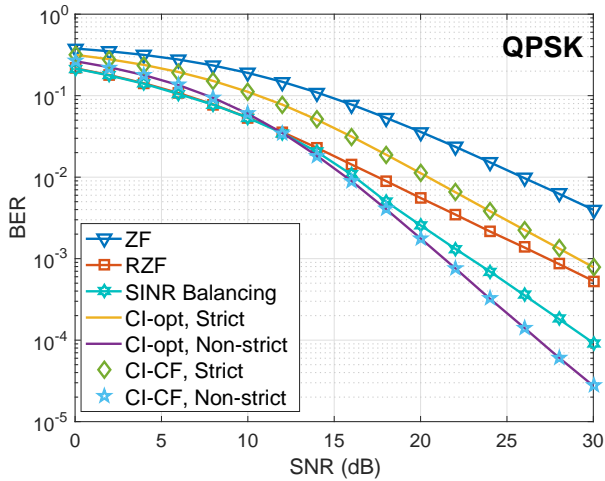}
\caption{BER v.s. transmit SNR, QPSK, $N_t=K=8$}
\end{figure} 

\begin{figure}[!t]
\centering
\includegraphics[scale=0.45]{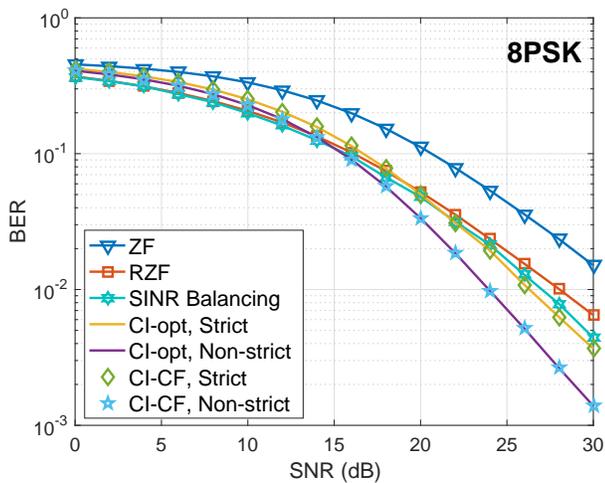}
\caption{BER v.s. transmit SNR, 8PSK, $N_t=K=8$}
\end{figure} 

In Fig. 4, we compare the bit error rate (BER) performance of different schemes with QPSK modulation, and $N_t=K=8$. As can be observed, the CI beamforming approaches for both strict phase rotation and non-strict phase rotation achieve an improved performance over the ZF approach, and the gain for non-strict phase rotation is more significant. For the CI beamforming for non-strict phase rotation at high SNR regime, we observe a SNR gain of more than 10dB over ZF and 8dB SNR gain over RZF. Moreover, we observe that the proposed iterative closed-form algorithm achieves exactly the same performance as the optimization-based CI beamforming, which validates the effectiveness of the proposed method in Section V.

\begin{figure}[!t]
\centering
\includegraphics[scale=0.45]{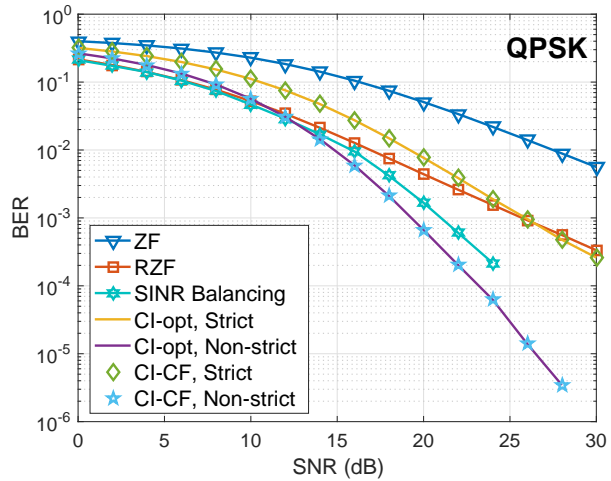}
\caption{BER v.s. transmit SNR, QPSK, $N_t=K=12$}
\end{figure} 

In Fig. 5, we show the BER performance with respect to the increasing transmit SNR when 8PSK modulation is employed, where $N_t=K=8$. Similarly, both transmit beamforming approaches based on the CI achieve an improved performance over the ZF method, and the proposed iterative closed-form schemes achieve the same performance as the optimization-based schemes. At high SNR ($\rho>20$dB), both CI-based approaches outperform the ZF-based schemes. For CI with non-strict phase rotation, we observe a SNR gain of over 7dB compared to ZF, and a SNR gain of 5dB compared to RZF precoding.

Fig. 6 shows the BER performance of different schemes for QPSK with $N_t=K=12$, where a similar BER trend can be observed. Particularly, comparing Fig. 4 and Fig. 6, we observe that the performance gains of the CI-based approaches over the conventional ZF precoding are more significant with the increasing number of antennas and users.

\begin{figure}[!t]
\centering
\includegraphics[scale=0.45]{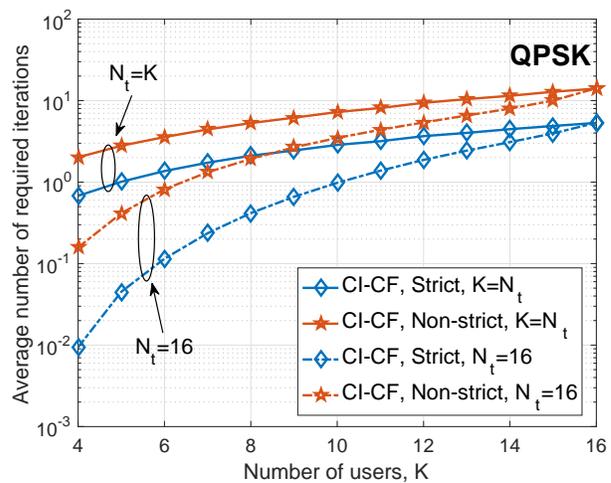}
\caption{Average number of iterations required for the iterative closed-form scheme, QPSK, $N_t=K$ and $N_t=16$}
\end{figure}

In Fig. 7, the average number of iterations required for the proposed iterative scheme is numerically studied with the increasing number of users, where we include two cases of $N_t=K$ and $N_t=16$. We observe that the average number of iterations increases with the increase in the number of users, as a larger number of users means a high possibility that more entries in $\bf a$ can be negative. The non-strict phase rotation requires more iterations than the strict phase rotation because the problem size is doubled. We also observe that when the number of users $K$ is small, the average number of iterations can be smaller than 1 because the number of iterations is zero when ${\cal S}=\emptyset$.

\begin{figure}[!t]
\centering
\includegraphics[scale=0.45]{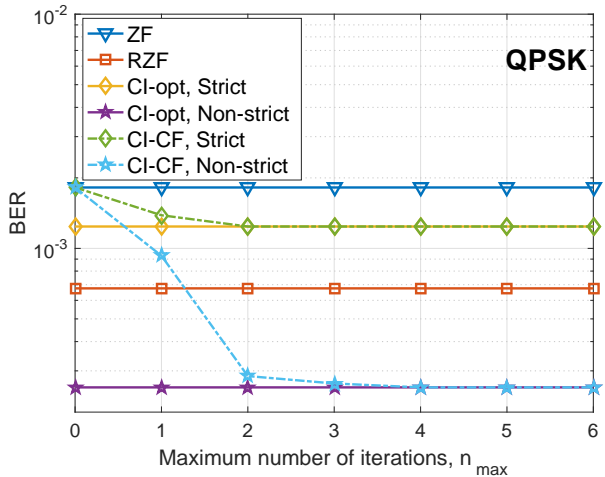}
\caption{BER v.s. maximum number of iterations $n_{\rm max}$, QPSK, $N_t=K=4$, SNR=30dB}
\end{figure}

To show the flexible performance-complexity tradeoff for the proposed algorithm, in Fig. 8 we depict the BER performance of the proposed iterative approach with respect to the maximum iteration number $n_{\rm max}$, where $N_t=K=4$. As expected, we observe that the performance of the iterative approach is identical to the conventional ZF approach when $n_{\rm max}=0$. With $n_{\rm max}$ increases, the performance of the iterative method approaches the optimal CI-based beamforming, which validates our statement in Section V-C.

\begin{figure}[!t]
\centering
\includegraphics[scale=0.45]{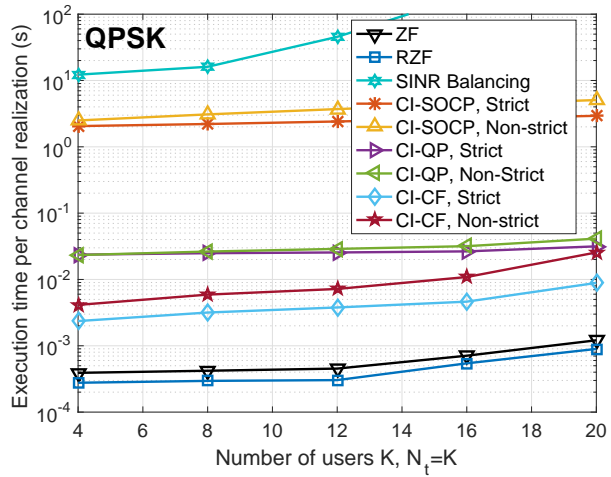}
\caption{Execution time required for different schemes, QPSK, $N_t=K$, 5000 channel realizations}
\end{figure} 

In Fig. 9, we compare the execution time required for each scheme to show the complexity benefits of the proposed iterative closed-form scheme over 5000 channel realizations, where `CI-QP, Strict/Non-Strict' refers to the QP optimizations ${\cal P}_4$ and ${\cal P}_8$, and `CI-SOCP, Strict/Non-Strict' refers to the SOCP optimizations ${\cal P}_1$ and ${\cal P}_5$. It is observed that the optimization for non-strict phase rotation requires more time to obtain the optimal solution that the strict-phase rotation because of the larger problem size. It is also observed that solving the equivalent QP optimization is much faster than solving the original SOCP optimization. More importantly, our proposed iterative scheme is more time-efficient than the QP algorithms, which motivates the use of the symbol-level CI beamforming in practice.

\section{Conclusion}
In this paper, we study the symbol-level downlink beamforming schemes based on CI, where both the strict and non-strict phase rotation cases are considered. By analyzing the optimization problems with Lagrangian and KKT conditions, we firstly obtain the optimal structure of the beamforming matrix, and further transform the optimization into a QP over a simplex by formulating the dual problem. We show that the optimizations for strict and non-strict phase rotation are equivalent in terms of the problem formulation. We further illustrate that ZF precoding is a special case and lower-bound of the CI beamforming. The proposed iterative closed-form scheme is shown to achieve an identical performance to the optimization-based schemes with a reduced computational cost, which enables the use of symbol-level CI beamforming in practical wireless systems.

\ifCLASSOPTIONcaptionsoff
  \newpage
\fi

\bibliographystyle{IEEEtran}
\bibliography{refs.bib}

\begin{thebibliography}{10}
\providecommand{\url}[1]{#1}
\csname url@samestyle\endcsname
\providecommand{\newblock}{\relax}
\providecommand{\bibinfo}[2]{#2}
\providecommand{\BIBentrySTDinterwordspacing}{\spaceskip=0pt\relax}
\providecommand{\BIBentryALTinterwordstretchfactor}{4}
\providecommand{\BIBentryALTinterwordspacing}{\spaceskip=\fontdimen2\font plus
\BIBentryALTinterwordstretchfactor\fontdimen3\font minus
  \fontdimen4\font\relax}
\providecommand{\BIBforeignlanguage}[2]{{%
\expandafter\ifx\csname l@#1\endcsname\relax
\typeout{** WARNING: IEEEtran.bst: No hyphenation pattern has been}%
\typeout{** loaded for the language `#1'. Using the pattern for}%
\typeout{** the default language instead.}%
\else
\language=\csname l@#1\endcsname
\fi
#2}}
\providecommand{\BIBdecl}{\relax}
\BIBdecl

\bibitem{r1}
L.~Zheng and D.~N.~C. Tse, ``{Diversity and Multiplexing: A Fundamental
  Tradeoff in Multiple-Antenna Channels},'' \emph{IEEE Trans. Inf. Theory},
  vol.~49, no.~5, pp. 1073--1096, May 2003.

\bibitem{r2}
M.~Costa, ``{Writing on Dirty Paper},'' \emph{IEEE Trans. Inf. Theory}, vol.
  IT-29, no.~3, pp. 439--441, May 1983.

\bibitem{r3}
L.~Sun and M.~Lei, ``{Quantized CSI-based Tomlinson-Harashima Precoding in
  Multiuser MIMO Systems},'' \emph{IEEE Trans. Wireless Commun.}, vol.~12,
  no.~3, pp. 1118--1126, Mar. 2013.

\bibitem{r4}
B.~M. Hochwald, C.~B. Peel, and A.~L. Swindlehurst, ``{A Vector-Perturbation
  Technique for Near-Capacity Multiantenna Multiuser Communication-part II:
  Perturbation},'' \emph{IEEE Trans. Commun.}, vol.~53, no.~3, pp. 537--544,
  Mar. 2005.

\bibitem{r5}
T.~Haustein, C.~von Helmolt, E.~Jorswieck, V.~Jungnickel, and V.~Pohl,
  ``{Performance of MIMO Systems with Channel Inversion},'' in \emph{Vehicular
  Technology Conference. IEEE 55th Vehicular Technology Conference. VTC Spring
  2002 (Cat. No.02CH37367)}, vol.~1, 2002, pp. 35--39.

\bibitem{r6}
C.~B. Peel, B.~M. Hochwald, and A.~L. Swindlehurst, ``{A Vector-Perturbation
  Technique for Near-Capacity Multiantenna Multiuser Communication-part I:
  Channel Inversion and Regularization},'' \emph{IEEE Trans. Commun.}, vol.~53,
  no.~1, pp. 195--202, Jan. 2005.

\bibitem{r7}
A.~Wiesel, Y.~C. Eldar, and S.~Shamai~(Shitz), ``{Linear Precoding via Conic
  Optimization for Fixed MIMO Receivers},'' \emph{IEEE Trans. Sig. Process.},
  vol.~54, no.~1, pp. 161--176, Jan. 2006.

\bibitem{r8}
M.~F. Hanif, L.-N. Tran, A.~Tolli, and M.~Juntti, ``{Computationally Efficient
  Robust Beamforming for SINR Balancing in Multicell Downlink with Applications
  to Large Antenna Array Systems},'' \emph{IEEE Trans. Commun.}, vol.~62,
  no.~6, pp. 1908--1920, June 2014.

\bibitem{r9}
F.~Wang, X.~Wang, and Y.~Zhu, ``{Transmit Beamforming for Multiuser Downlink
  with Per-Antenna Power Constraints},'' in \emph{2014 IEEE International
  Conference on Communications (ICC)}, Sydney, NSW, 2014, pp. 4692--4697.

\bibitem{r10}
M.~Schubert and H.~Boche, ``{Solution of the Multiuser Downlink Beamforming
  Problem with Individual SINR Constraints},'' \emph{IEEE Trans. Veh. Tech.},
  vol.~53, no.~1, pp. 18--28, Jan. 2004.

\bibitem{r11}
M.~Bengtsson and B.~Ottersten, ``{Optimal and Suboptimal Transmit
  Beamforming},'' \emph{Handbook of Antennas in Wireless Communications}, Jan.
  2001.

\bibitem{r12}
N.~D. Sidiropoulos, T.~N. Davidson, and Z.-Q. Luo, ``{Transmit Beamforming for
  Physical-Layer Multicasting},'' \emph{IEEE Trans. Sig. Process.}, vol.~54,
  no.~6, pp. 2239--2251, June 2006.

\bibitem{r13}
E.~Karipidis, N.~D. Sidiropoulos, and Z.-Q. Luo, ``{Quality of Service and
  Max-Min Fair Transmit Beamforming to Multiple Cochannel Multicast Groups},''
  \emph{IEEE Trans. Sig. Process.}, vol.~56, no.~3, pp. 1268--1279, Mar. 2008.

\bibitem{r14}
C.~Masouros, T.~Ratnarajah, M.~Sellathurai, C.~B. Papadias, and A.~K. Shukla,
  ``{Known Interference in the Cellular Downlink: A Performance Limiting Factor
  or a Source of Green Signal Power?}'' \emph{IEEE Commun. Mag.}, vol.~51,
  no.~10, pp. 162--171, Oct. 2013.

\bibitem{r15}
G.~Zheng, I.~Krikidis, C.~Masouros, S.~Timotheou, D.~A. Toumpakaris, and
  Z.~Ding, ``{Rethinking the Role of Interference in Wireless Networks},''
  \emph{IEEE Commun. Mag.}, vol.~52, no.~11, pp. 152--158, Nov. 2014.

\bibitem{r30}
C.~Masouros and E.~Alsusa, ``{A Novel Transmitter-Based Selective-Precoding
  Technique for DS/CDMA Systems},'' \emph{IEEE Sig. Process. Lett.}, vol.~14,
  no.~9, pp. 637--640, Sept. 2007.

\bibitem{r16}
------, ``{Dynamic Linear Precoding for the Exploitation of Known Interference
  in MIMO Broadcast Systems},'' \emph{IEEE Trans. Wireless Commun.}, vol.~8,
  no.~3, pp. 1396--1401, Mar. 2009.

\bibitem{r17}
C.~Masouros, ``{Correlation Rotation Linear Precoding for MIMO Broadcast
  Communicaitons},'' \emph{IEEE Trans. Sig. Process.}, vol.~59, no.~1, pp.
  252--262, Jan. 2011.

\bibitem{r18}
C.~Masouros and G.~Zheng, ``{Exploiting Known Interference as Green Signal
  Power for Downlink Beamforming Optimization},'' \emph{IEEE Trans. Sig.
  Process.}, vol.~63, no.~14, pp. 3628--3640, July 2015.

\bibitem{r19}
M.~Alodeh, S.~Chatzinotas, and B.~Ottersten, ``{Constructive Multiuser
  Interference in Symbol Level Precoding for the MISO Downlink Channel},''
  \emph{IEEE Trans. Sig. Process.}, vol.~63, no.~9, pp. 2239--2252, May 2015.

\bibitem{r20}
------, ``{Energy-Efficient Symbol-Level Precoding in Multiuser MISO based on
  Relaxed Detection Region},'' \emph{IEEE Trans. Wireless Commun.}, vol.~15,
  no.~5, pp. 3755--3767, May 2016.

\bibitem{r31}
C.~Masouros, M.~Sellathurai, and T.~Ratnarajah, ``{Vector Perturbation based on
  Symbol Scaling for Limited Feedback MISO Downlinks},'' \emph{IEEE Trans. Sig.
  Process.}, vol.~62, no.~3, pp. 562--571, Feb. 2014.

\bibitem{r21}
M.~Alodeh, S.~Chatzinotas, and B.~Ottersten, ``{Symbol-Level Multiuser MISO
  Precoding for Multi-Level Adaptive Modulation},'' \emph{IEEE Trans. Wireless
  Commun.}, vol.~16, no.~8, pp. 5511--5524, Aug. 2017.

\bibitem{r22}
C.~Masouros and T.~Ratnarajah, ``{Interference as a Source of Green Signal
  Power in Cognitive Relay Assisted Co-Existing MIMO Wireless Transmissions},''
  \emph{IEEE Trans. Commun.}, vol.~60, no.~2, pp. 525--536, Feb. 2012.

\bibitem{r23}
K.~L. Law, C.~Masouros, and M.~Pesavento, ``{Transmit Precoding for
  Interference Exploitation in the Underlay Cognitive Radio Z-Channel},''
  \emph{IEEE Trans. Sig. Process.}, vol.~65, no.~14, pp. 3617--3631, July 2017.

\bibitem{r24}
P.~V. Amadori and C.~Masouros, ``{Constant Envelope Precoding by Interference
  Exploitation in Phase Shift Keying-Modulated Multiuser Transmission},''
  \emph{IEEE Trans. Wireless Commun.}, vol.~16, no.~1, pp. 538--550, Jan. 2017.

\bibitem{r25}
S.~Timotheou, G.~Zheng, C.~Masouros, and I.~Krikidis, ``{Exploiting
  Constructive Interference for Simultaneous Wireless Information and Power
  Transfer in Multiuser Downlink Systems},'' \emph{IEEE J. Sel. Areas Commun.},
  vol.~34, no.~5, pp. 1772--1784, May 2016.

\bibitem{r26}
A.~Li and C.~Masouros, ``{Exploiting Constructive Mutual Coupling in P2P MIMO
  by Analog-Digital Phase Alignment},'' \emph{IEEE Trans. Wireless Commun.},
  vol.~16, no.~3, pp. 1948--1962, Mar. 2017.

\bibitem{r27}
L.~Vandenberghe and S.~Boyd, \emph{{Convex Optimization}}.\hskip 1em plus 0.5em
  minus 0.4em\relax Cambridge University Press, 2004.

\bibitem{r28}
P.~Wolfe, ``{The Simplex Method for Quadratic Programming},''
  \emph{Econometrica}, vol.~27, no.~3, pp. 382--398, July 1959.

\bibitem{r29}
G.~Cornuejols and R.~Tutuncu, \emph{{Optimization Methods in Finance}}.\hskip
  1em plus 0.5em minus 0.4em\relax Cambridge University Press, Dec. 2006.

\end{thebibliography}

\end{document}